\documentclass[
	aps, prd, reprint, a4paper,
	amsmath, amssymb, amsfonts, eqsecnum,
	superscriptaddress, showpacs, showkeys,
	nofootinbib 
]{revtex4-1}

\usepackage{ifthen}
\newboolean{prd}
\setboolean{prd}{false}
\newboolean{arxiv}
\setboolean{arxiv}{true}
\newboolean{notprd}
\setboolean{notprd}{true}
\ifprd
\setboolean{notprd}{false}
\fi
\newboolean{notarxiv}
\setboolean{notarxiv}{true}
\ifarxiv
\setboolean{notarxiv}{false}
\fi

\ifarxiv
\usepackage{graphics}
\else
\usepackage{CJK}
\fi

\ifnotarxiv
\ifnotprd
\usepackage{xcolor}
\xdefinecolor{mylinkcolor}{rgb}{0,0,0.5}
\usepackage[
	bookmarksnumbered, bookmarksopen, bookmarksopenlevel=2,
	breaklinks=true, colorlinks=true,
	filecolor=mylinkcolor, citecolor=mylinkcolor, linkcolor=mylinkcolor,
	urlcolor=mylinkcolor, menucolor=mylinkcolor,
]{hyperref}
\else
\newcommand{\texorpdfstring}[2]{{#1}}
\fi
\else
\usepackage{xcolor}
\xdefinecolor{mylinkcolor}{rgb}{0,0,0.5}
\usepackage[
	bookmarksnumbered, bookmarksopen, bookmarksopenlevel=1,
	colorlinks=true,
	filecolor=mylinkcolor, citecolor=mylinkcolor, linkcolor=mylinkcolor,
	urlcolor=mylinkcolor, menucolor=mylinkcolor,
]{hyperref}
\fi

\def\vct#1{\mathbf{#1}}

\def\nl{\\ & \quad}
\def\nlq{\\ & \quad \qquad}
\def\nnl{\nonumber \\ & \quad}
\def\nnlq{\nonumber \\ & \quad \qquad}

\def\pa{\partial}

\def\dd{d}

\def\htt{h^{\rm TT}}
\def\htts{h^{{\rm TT\, spin}}}

\def\intxp{\int d^3x^{\prime}}
\def\intx{\int d^3x}
\def\suma{\sum_a}
\def\sumb{\sum_{b \ne a}}

\def\sumap{\sum_{a^\prime}}
\def\sumbp{\sum_{b^\prime \ne a^\prime}}

\def\mapr{m_{a^\prime}}

\def\raap{r_{{a}{a^\prime}}}

\def\naap{n_{{a}{a^\prime}}}

\def\primevarsc{{a \rightarrow a^{\prime}}}
\newcommand{\primevar}[1]{#1^{\primevarsc}}

\def\dtex{\pa_t^{\text{ex}}}

\allowdisplaybreaks

\begin{document}

\ifnotarxiv
\begin{CJK*}{UTF8}{gbsn}
\fi

\title{Leading-order spin-orbit and spin(1)-spin(2) radiation-reaction Hamiltonians}

\ifnotarxiv
\author{Han Wang (王涵)}
\else
\author{Han Wang (\includegraphics{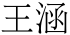})}
\fi
\affiliation{Theoretisch--Physikalisches Institut,
	Friedrich--Schiller--Universit\"at,
	Max--Wien--Platz 1, 07743 Jena, Germany, EU}

\author{Jan Steinhoff}
\email{jan.steinhoff@ist.utl.pt}
\affiliation{Theoretisch--Physikalisches Institut,
	Friedrich--Schiller--Universit\"at,
	Max--Wien--Platz 1, 07743 Jena, Germany, EU}
\affiliation{Centro Multidisciplinar de Astrof\'isica --- CENTRA, Departamento de F\'isica,
	Instituto Superior T\'ecnico --- IST, Universidade T\'ecnica de Lisboa, 
	Avenida Rovisco Pais 1, 1049-001 Lisboa, Portugal, EU}

\ifnotarxiv
\author{Jing Zeng (曾靓)}
\else
\author{Jing Zeng (\includegraphics{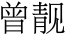})}
\fi
\email{jzeng@shao.ac.cn}
\affiliation{Key Laboratory for Research in Galaxies and Cosmology, Shanghai Astronomical Observatory, Nandan Road 80, Shanghai, 200030, China}

\author{Gerhard Sch\"afer}
\email{Gerhard.Schaefer@uni-jena.de}
\affiliation{Theoretisch--Physikalisches Institut,
	Friedrich--Schiller--Universit\"at,
	Max--Wien--Platz 1, 07743 Jena, Germany, EU}

\date{\today}

\begin{abstract}
In the present paper, the leading-order post-Newtonian spin-orbit and
spin(1)-spin(2) \emph{radiation-reaction} Hamiltonians are calculated. We utilize the canonical
formalism of Arnowitt, Deser, and Misner (ADM), which has shown to be valuable for this kind of calculation. The
results are valid for arbitrary many objects. The
energy loss is then computed and compared to well-known results for the energy flux
as a check.
\end{abstract}

\pacs{04.25.Nx, 04.20.Fy, 04.25.-g, 97.80.-d}
\keywords{post-Newtonian approximation; ADM canonical formalism; Gravitational radiation reaction; Binaries and multiple stars; Spinning bodies}

\maketitle

\ifnotarxiv
\end{CJK*}
\fi

\section{Introduction\label{sec:intro}}
Gravitational radiation reaction is a problem of great interest in the detection
of gravitational waves. For second and third generations of gravitational wave
detectors, a leading candidate source is the radiation-reaction induced inspiral and merger
of two compact objects like black holes or
neutron stars. Moreover, the effects of spins are important for the
emission of gravitational waves from such systems. Thus in order to develop
highly accurate theoretical templates for  gravitational wave detectors, one
must study the gravitational radiation reaction from compact binary systems with spin
effects.

In the present paper, the leading-order post-Newtonian (PN) spin-orbit and
spin(1)-spin(2) radiation-reaction, i.e., \emph{dissipative}, Hamiltonians are
calculated. This is the continuation of previous work in
\cite{Steinhoff:Wang:2009}, where the formalism was prepared, and also extends
the calculation of the 3.5PN point-mass Hamiltonian in
\cite{Jaranowski:Schafer:1997} to that of spinning objects. The contributions of the
spin-dependent Hamiltonians derived in the present paper to the equations of
motion are 2.5PN orders weaker than the corresponding leading-order
\emph{conservative} ones. Recently, the contributions to the motion of spinning
objects have just come within reach of experimental verifications
\cite{Everitt:others:2011, Breton:others:2008}. A further increase in precision of
experimental tests of general relativity will become available by creation and
subsequent improvement of gravitational wave astronomy
\cite{Pitkin:Reid:Rowan:Hough:2011, Sathyaprakash:Schutz:2009} in the future. 
For compact binary systems detectable by gravitational wave detectors, the Hamiltonians
derived in the present paper become relevant in the late inspiral phase 
if one or more of the binary's constituents is rapidly rotating. And rapidly
rotating black holes have been proved to be astrophysically realistic
\cite{McClintock:others:2011}. Therefore, the derivation of the Hamiltonians with spin effects is necessary for the detection of gravitational waves.

In this paper, we utilize the canonical formalism of Arnowitt, Deser, and Misner (ADM),
which has not only shown to be valuable for calculating the conservative dynamics within the
post-Newtonian and post-Minkowskian approximations (see,
e.g., \cite{Jaranowski:Schafer:1998, Damour:Jaranowski:Schafer:2001,
Ledvinka:Schafer:Bicak:2008}) but also for the dissipative part of the dynamics
\cite{Jaranowski:Schafer:1997} (with misprints corrected in
\cite{Konigsdorffer:Faye:Schafer:2003}). Notice that the ADM formalism was
extended from point-masses to objects with spins only recently
\cite{Steinhoff:Schafer:2009:2} (see also \cite{Steinhoff:2011,
Steinhoff:Schafer:Hergt:2008, Steinhoff:Wang:2009}). This extension is valid to
linear order in the single spins of the objects, which not only includes
spin-orbit but also spin(1)-spin(2) interactions. The remarkable structure of
the extended ADM formalism  of the inclusion of the matter into the canonical
field momentum [see Eq.\ (\ref{pican})] is passing an excellent test in the
present paper. For Hamiltonians of higher orders in spins see
\cite{Hergt:Schafer:2008:2, Hergt:Schafer:2008, Steinhoff:Hergt:Schafer:2008:1,
Hergt:Steinhoff:Schafer:2010:1, Steinhoff:2011}.

Energy and angular momentum flux relevant for the PN order in question has been well known 
(see \cite{Kidder:1995}, for the next-to-leading-order calculation see
\cite{Blanchet:Buonanno:Faye:2006, *Blanchet:Buonanno:Faye:2006:err,
*Blanchet:Buonanno:Faye:2006:err:2, Porto:Ross:Rothstein:2010}). Based on these
results, secular equations of motion for the orbital elements corresponding to
the leading-order spin-orbit and spin(1)-spin(2) radiation-reaction equations of
motion were obtained in \cite{Gergely:Perjes:Vasuth:1998, Gergely:1999,
Gergely:2000}. The general equations of motion at this order were calculated in
\cite{Will:2005, Wang:Will:2007, Zeng:Will:2007} within the harmonic gauge.
The Hamiltonians calculated in the present paper provide a compact expression
which contains these general equations of motion (but within a different gauge).
And most importantly, the results in the present paper are valid for arbitrary many
object systems. The derived Hamiltonians are then applied to the calculation of the energy
loss of a binary system, which is then compared with the well-known energy flux as a check.

The conservative leading-order (PN) spin interactions for self-gravitating objects
were derived some time ago \cite{Barker:OConnell:1975, DEath:1975,
Barker:OConnell:1979, Thorne:Hartle:1985}, see also \cite{Thorne:1980,
Poisson:1998}. For the
leading-order spin(1)-spin(1) radiation-reaction level calculations see,
e.g., \cite{Poisson:1998, Gergely:1999, Gergely:Keresztes:2003}. However, only recently the conservative next-to-leading-order spin effects
could be treated, starting with the spin-orbit equations of motion in harmonic
gauge \cite{Tagoshi:Ohashi:Owen:2001} (with some extensions and misprints corrected in
\cite{Faye:Blanchet:Buonanno:2006}). A corresponding conservative Hamiltonian in
the ADM gauge was obtained in \cite{Damour:Jaranowski:Schafer:2008:1}. The
complete next-to-leading-order spin(1)-spin(2) conservative Hamiltonian was
first given in \cite{Steinhoff:Hergt:Schafer:2008:2}. Other derivations of the
conservative next-to-leading-order spin-orbit and spin(1)-spin(2) dynamics can
be found in \cite{Perrodin:2010, Porto:2010, Levi:2010, Porto:Rothstein:2008:1,
*Porto:Rothstein:2008:1:err, Levi:2008} and a generalization to arbitrary many
objects succeeded in \cite{Hartung:Steinhoff:2010}. Notice that the results
given in the present paper are already valid for arbitrary many objects. Also
the conservative next-to-leading-order spin(1)-spin(1) interaction of black hole
and/or neutron star binaries was derived recently
\cite{Steinhoff:Hergt:Schafer:2008:1, Hergt:Schafer:2008,
Porto:Rothstein:2008:2, *Porto:Rothstein:2008:2:err, Steinhoff:Schafer:2009:1,
Hergt:Steinhoff:Schafer:2010:1}. The latter requires a modeling of the
spin-induced quadrupole deformation, see \cite{Poisson:1998,
Laarakkers:Poisson:1999}. Very recently, the conservative spin-dependent part of
the post-Newtonian Hamiltonian was extended even to next-to-next-to-leading
order for both the spin-orbit \cite{Hartung:Steinhoff:2011:1} and the
spin(1)-spin(2) \cite{Hartung:Steinhoff:2011:2} cases. A potential for the
spin(1)-spin(2) case was simultaneously calculated within an effective
field theory approach \cite{Levi:2011}. Notice that the conservative
next-to-next-to-leading-order spin(1)-spin(2) Hamiltonian and the spin-orbit
radiation-reaction Hamiltonian derived in the present paper are both of the
order 4PN for maximally rotating objects. However, not all spin-dependent
Hamiltonians up to 4PN for maximally rotating objects are known yet. We will
in most cases use the phrase "formal n-th PN order" to represent our counting of
PN orders in the present paper. This gives PN orders different from the
maximally rotating case, which we also occasionally refer to in the present
paper (for a more detailed discussion see, e.g., Appendix A of
\cite{Steinhoff:Wang:2009}). But one should be aware that the spins are in fact
further (independent) expansion variables. Spin effects were also considered within the
post-Minkowskian approximation \cite{Goenner:Gralewski:Westpfahl:1967,
Bennewitz:Westpfahl:1971}. 

The paper is organized as follows. First the ADM formalism is reviewed in
Sec.\ \ref{sec:formalism}. Then formal expressions for the radiation-reaction
Hamiltonians in question are derived in Sec.\ \ref{sec:rrhamiltonian}. Integrals
appearing in these formal expressions are performed in
Sec.\ \ref{sec:calculation}. In Sec.\ \ref{sec:energyloss}, the derived
Hamiltonians are applied to the calculation of the energy loss, which is then
compared with the energy flux. Finally, conclusions are given in
Sec.\ \ref{sec:conclusions}.

Our units are such that $c=1$, but for the Newtonian gravitational constant $G$
no convention will be used. This allows an easy transition to the different
conventions for $G$ used in \cite{Jaranowski:Schafer:1997} and
\cite{Steinhoff:Wang:2009}. For the signature of spacetime, we choose +2. Latin
indices from the beginning of the alphabet, such as a, b, label the individual objects. Greek
indices run over $0,1,2,3$. Latin indices from the middle of the alphabet run over
$1,2,3$. Round brackets around an index denote a local basis, while round
brackets around a number denote the formal order in $c^{-1}$, as in
\cite{Steinhoff:Wang:2009, Steinhoff:Schafer:Hergt:2008}. A 3-vector $x^i$ is also denoted by $\vct{x}$.
Square brackets denote index antisymmetrization and round brackets index
symmetrization, i.e.,
$a^{( \mu} b^{\nu )} = \frac{1}{2} (a^{\mu} b^{\nu} + a^{\nu} b^{\mu})$.

\section{The ADM formalism\label{sec:formalism}}
In this section, we provide a short overview of the ADM canonical formalism after
gauge fixing \cite{Arnowitt:Deser:Misner:1962, *Arnowitt:Deser:Misner:2008}, see also
\cite{Regge:Teitelboim:1974, DeWitt:1967}. The Hamiltonian is given by the
ADM energy expressed in terms of certain canonical variables, which also
requires a (at least approximate) solution of the field constraints.

The constraints of the gravitational field read
\begin{gather}
\frac{1}{16\pi G \sqrt{\gamma}} \left[ \gamma \text{R}
	+ \frac{1}{2} \left( \gamma_{ij} \pi^{ij} \right)^2
	- \gamma_{ij} \gamma_{k l} \pi^{ik} \pi^{jl}\right]
	= \mathcal{H}^{\text{matter}} \,, \label{ham} \\
- \frac{1}{8\pi G} \gamma_{ij} \pi^{jk}_{~~ ; k} = \mathcal{H}^{\text{matter}}_i \,, \label{mom}
\end{gather}
with the definitions
\begin{align}
\pi^{ij} &= - \sqrt{\gamma} (\gamma^{ik}\gamma^{jl} - \gamma^{ij}\gamma^{kl})K_{kl} \,, \\
\mathcal{H}^{\rm matter} &= \sqrt{\gamma}T_{\mu\nu} n^{\mu}n^{\nu} \,, \\
\mathcal{H}^{\rm matter}_i &= - \sqrt{\gamma}T_{i \nu} n^{\nu} \,.
\end{align}
They arise as certain projections of the Einstein field equations with respect to a
timelike unit 4-vector $n_{\mu}$ with components $n_{\mu} = (-N, 0, 0, 0)$ or
$n^{\mu} = (1, -N^i) / N$. Here, $\gamma_{ij}$ is the induced three-dimensional
metric of the hypersurfaces orthogonal to $n_{\mu}$, $\gamma$ its determinant,
$\text{R}$ the three-dimensional Ricci scalar,
$K_{ij} = - ( \tfrac{1}{2} \gamma_{ij,0} - N_{(i;j)} ) / N$ the extrinsic
curvature, $N$ the lapse function, $N^i$ the shift vector,
$ \sqrt{\gamma}T_{\mu\nu}$ the stress-energy tensor density of the matter
system, and semicolon denotes the three-dimensional covariant derivative. Partial
coordinate derivatives $\pa_i$ are also indicated by commas.

For nonspinning objects, $\frac{1}{16\pi G} \pi^{ij}$ is the canonical momentum
conjugate to $\gamma_{ij}$ before gauge fixing. For spinning objects, the
canonical field momentum has to be adapted, see \cite{Steinhoff:Schafer:2009:2,
Steinhoff:Wang:2009}. We write
\begin{align}\label{pican}
	\pi^{ij}_{\text{can}} &= \pi^{ij} + \pi^{ij}_{\text{matter}} \,,
\end{align}
where $\pi^{ij}_{\text{matter}}$ contains spin-corrections. Throughout this
paper we use the ADM transverse-traceless (TT) gauge, which is defined by:
\begin{gather}
\partial_j ( \gamma_{ij} - \tfrac{1}{3} \delta_{ij} \gamma_{kk} ) = 0 \,, \label{ADMTTg} \\
\pi^{ii}_{\text{can}} = 0 \,. \label{ADMTTpi}
\end{gather}
Here, $\delta_{ij}$ is the Kronecker delta. And one has the decompositions:
\begin{align}
	\gamma_{ij} &= \left( 1 + \frac{\phi}{8} \right)^4 \delta_{ij} + h^{\text{TT}}_{ij} \,,
		\label{gdecomp} \\
	\pi^{ij}_{\text{can}} &= \pi^{ij\text{TT}}_{\text{can}} + \tilde{\pi}^{ij}_{\text{can}} \,.
\end{align}
Notice that $i$, $j$, and $k$, etc., run over 1, 2, 3, 
and upper or lower an index is from now on done with the flat metric, thus changes nothing in the equations. 
We will ignore the difference of upper and lower indexes later and any two identical indexes can contract with no need to be one upper and one lower. 
$h^{\text{TT}}_{ij}$ and $\pi^{ij\text{TT}}_{\text{can}}$ are transverse-traceless,
e.g., $h^{\text{TT}}_{ii} = h^{\text{TT}}_{ij,j}=0$, and $\tilde{\pi}^{ij}_{\text{can}}$ is
related to the vector potentials $V^i_{\text{can}}$ and $\tilde{\pi}^i_{\text{can}}$ by:
\begin{align}
\tilde{\pi}^{ij}_{\text{can}} &= V^i_{\text{can}, j}
	+ V^j_{\text{can}, i} - \frac{2}{3} \delta_{ij} V^k_{\text{can}, k} \,, \\
	&= \tilde{\pi}^i_{\text{can}, j} + \tilde{\pi}^j_{\text{can}, i}
	- \frac{1}{2} \delta_{ij} \tilde{\pi}^k_{\text{can}, k}
	- \frac{1}{2} \Delta^{-1} \tilde{\pi}^k_{\text{can}, ijk} \,.
\end{align}
It holds that:
\begin{align}
V^i_{\text{can}} &= \left( \delta_{ij} - \frac{1}{4} \partial_i \partial_j \Delta^{-1} \right) \tilde{\pi}^j_{\text{can}} \,, \\ \tilde{\pi}^i_{\text{can}} &= \Delta^{-1} \pi^{ij}_{\text{can},j}
	= \Delta^{-1} \tilde{\pi}^{ij}_{\text{can},j} \,, \\
\pi^{ij\text{TT}}_{\text{can}} &= \delta^{\text{TT}ij}_{kl} \pi^{kl}_{\text{can}} \,,
\end{align}
with the inverse Laplacian $\Delta^{-1}$ and
\begin{equation}\label{TTproj}
\begin{split}
\delta^{\text{TT}kl}_{ij} &= \tfrac{1}{2} [(\delta_{il}-\Delta^{-1}\pa_{i}\pa_{l})(\delta_{jk}-\Delta^{-1}\pa_{j}\pa_{k}) \nl
+(\delta_{ik}-\Delta^{-1}\pa_{i}\pa_{k})(\delta_{jl}-\Delta^{-1}\pa_{j}\pa_{l}) \nl -(\delta_{kl}-\Delta^{-1}\pa_{k}\pa_{l})(\delta_{ij}-\Delta^{-1}\pa_{i}\pa_{j})] \,.
\end{split}
\end{equation}
The canonical field variables after gauge fixing are $h^{\text{TT}}_{ij}$ and
$\pi^{ij\text{TT}}_{\text{can}}$.

In order to obtain the ADM Hamiltonian, the four field constraints must be
solved for the four variables $\phi$ and $\tilde{\pi}^i_{\text{can}}$ in terms of
$h^{\text{TT}}_{ij}$, $\pi^{ij\text{TT}}_{\text{can}}$ and canonical matter
variables, which enter through the stress-energy tensor via the source terms
$\mathcal{H}^{\rm matter}$ and $\mathcal{H}^{\rm matter}_i$ (for the linear
order source terms in spin, see \cite{Steinhoff:Schafer:2009:2,Steinhoff:Wang:2009,
Steinhoff:2011} and also \cite{Steinhoff:Schafer:Hergt:2008}). The canonical
matter variables are the canonical position $\hat{z}_a^i$, momentum $P_{a i}$,
and spin-tensor $S_{a (i) (j)}$ of the $a$-th object. An analytic solution for
$\phi$ and $\tilde{\pi}^i_{\text{can}}$, however, can in general only be given in some
approximation scheme. The ADM Hamiltonian is then given by:
\begin{equation}\label{HADM}
H_{\text{ADM}} = - \frac{1}{16\pi G} \int \dd^3x \, \Delta
	\phi [ \hat{z}^i_a, P_{a i}, S_{a(i)}, \htt_{ij}, \pi^{ij\text{TT}}_{\text{can}} ] \,,
\end{equation}
where $S_{a(i)} = \frac{1}{2} \epsilon_{ijk} S_{a(j)(k)}$ and $\epsilon_{ijk}$
is the completely antisymmetric Levi--Civita symbol. $H_{\text{ADM}}$ is the ADM
energy expressed in terms of the canonical variables mentioned above. The
Poisson brackets read
\begin{align}
\{ h^{\text{TT}}_{ij}({\bf x}), \pi^{kl\text{TT}}_{\text{can}}({\bf x}') \}
	&= 16\pi G \delta^{\text{TT}kl}_{ij}\delta({\bf x} - {\bf x}') \,, \label{fieldPB} \\
\{ \hat{z}^i_a, P_{a j} \} &= \delta_{ij} \,, \\
\{ S_{a(i)}, S_{a(j)} \} &= \epsilon_{ijk} S_{a(k)} \,,
\end{align}
all others are zero.

\section{Radiation-reaction Hamiltonians up to formal 3.5PN level\label{sec:rrhamiltonian}}
In this section, we generalize the derivation of the radiation-reaction
Hamiltonians up to the formal 3.5PN level performed in
\cite{Jaranowski:Schafer:1997} so it becomes applicable to the spinning case.

\subsection{Interaction Hamiltonian and wave equation}
We split the ADM Hamiltonian $H_{\text{ADM}}$ into matter, field, and
interaction parts, i.e.,
\begin{equation}
H_{\text{ADM}} = H^{\text{matter}} + H^{\text{field}} + H^{\text{int}} \,,
\end{equation}
where the matter part $H^{\text{matter}}$ is independent of the (truly
dynamical) canonical field variables $h^{\text{TT}}_{i j}$ and
$\pi^{i j \text{TT}}_{\text{can}}$, the field part $H^{\text{field}}$ is
independent of the canonical matter variables and reads explicitly:
\begin{equation}\label{Hfield}
H^{\text{field}} = \frac{1}{16\pi G} \intx \left[ \frac{1}{4}(h^{\rm TT}_{ij,k})^2 +(\pi^{i j \text{TT}}_{\text{can}})^2 \right] \,,
\end{equation}
and the interaction part $H^{\text{int}}$ depends on both canonical matter and
field variables. The interaction Hamiltonian up to and including the formal
3.5PN level reads (\cite{Steinhoff:Wang:2009}, see also \cite{Jaranowski:Schafer:1997})
\begin{align}
H^{\text{int}} &= \frac{1}{16\pi G} \int \dd^3 x \, \bigg[
	\left( B_{(4)ij} + \hat{B}_{(6)ij} \right) h^{\text{TT}}_{i j} \nnlq
	- 2\pi G \mathcal{H}^{\rm matter}_{(2)} \left(h^{\text{TT}}_{i j}\right)^2
	- \frac{1}{4} \phi_{(2)} \left(h^{\text{TT}}_{i j , k}\right)^2 \nnlq
	+ 2 ( V^i_{(3)} \phi_{(2),j} - \pi^{i j}_{(5)\text{matter}} ) \pi^{i j \text{TT}}_{\text{can}}
	\bigg] \,, \label{Hint}
\end{align}
where
\begin{align}\label{B4ham}
B_{(4)ij} &= 16\pi G \frac{\delta \left(
		\int{ d^3 x \, \mathcal{H}^{\rm matter}_{(8)}  } \right)}
		{\delta h^{\text{TT}}_{i j}}
	- \frac{1}{8} \phi_{(2) , i} \phi_{(2) , j} \,,
\end{align}
and $V^i_{(3)}$ is a field quantity which will be discussed in Sec. \ref{sec:calculation}. 
$\hat{B}_{(6)ij}$ is given by a similar expression [see (5.14) in
\cite{Steinhoff:Wang:2009}]. For comparison with \cite{Jaranowski:Schafer:1997}, notice that
$2 \delta^{\text{TT}ij}_{kl} ( V^k_{(3)} \phi_{(2),l} ) = - \delta^{\text{TT}ij}_{kl} 
( \phi_{(2)} \tilde{\pi}^{kl}_{(3)} )$ [$\tilde{\pi}^{kl}_{(3)}$ is another field quantity which will be discussed later].
Further, in \cite{Jaranowski:Schafer:1997} the quantity $A_{(4)ij} = 2 B_{(4)ij}$ is used in this paper.

The equations of motion for the canonical field variables follow from the ADM
Hamiltonian by virtue of the Poisson brackets (\ref{fieldPB}) as:
\begin{align}
\frac{1}{16\pi G} \dot{h}^{\text{TT}}_{i j} &=
	\delta^{\text{TT} i j}_{k l} \frac{\delta H_{\text{ADM}}}{\delta \pi^{k l \text{TT}}_{\text{can}}} \,, \label{httdot} \\
\frac{1}{16\pi G} \dot{\pi}^{i j \text{TT}}_{\text{can}} &=
	- \delta^{\text{TT} i j}_{k l} \frac{\delta H_{\text{ADM}}}{\delta h^{\text{TT}}_{k l}} \,. \label{pittdot}
\end{align}
Here the dot over a variable denotes the partial time derivative
$\pa_t \equiv \frac{\pa}{\pa t}$. For quantities not depending on the
hypersurface coordinate $\vct{x}$, this is to be understood as the ordinary
time derivative. In terms of the interaction Hamiltonian $H^{\text{int}}$, the
field equations read
\begin{align}
\frac{1}{16\pi G} \Box h^{\text{TT}}_{i j} &= \delta^{\text{TT} i j}_{k l}
		\left[ 2 \frac{\delta H^{\text{int}}}{\delta h^{\text{TT}}_{k l}}
	- \frac{\partial}{\partial t} \frac{\delta H^{\text{int}}}{\delta \pi^{k l \text{TT}}_{\text{can}}} \right] \,, \label{boxhtt} \\
\frac{1}{16\pi G} \pi^{i j \text{TT}}_{\text{can}} &= \frac{1}{2} \left[ \frac{1}{16\pi G} \dot{h}^{\text{TT}}_{i j}
	- \delta^{\text{TT} i j}_{k l} \frac{\delta H^{\text{int}}}{\delta \pi^{k l \text{TT}}_{\text{can}}} \right] \,,
\end{align}
with $\Box = \Delta - \pa_t^2$. To arrive at these expressions, the explicit form
of $H^{\text{field}}$ is used as in Eq.\ (\ref{Hfield}). Notice that it is easier to
implement the boundary condition of no incoming gravitational radiation for a
wave equation like Eq.\ (\ref{boxhtt}) than for a system of first-order differential
equations like Eqs.\ (\ref{httdot}) and (\ref{pittdot}). Inserting the 3.5PN-accurate
interaction Hamiltonian [Eq.\ (\ref{Hint})], one gets
\begin{align}
\Box h^{\text{TT}}_{i j} &= \delta^{\text{TT} i j}_{k l}
	\bigg[ 2 B_{(4)kl} + 2 B_{(6)kl}
	- 8\pi G \mathcal{H}^{\rm matter}_{(2)} h^{\text{TT}}_{kl} \nnl
	+ \left( \phi_{(2)} h^{\text{TT}}_{kl,m} \right)_{,m}
	- 2 \frac{\pa}{\pa t} \left( V^k_{(3)} \phi_{(2),l} \right) \bigg] \,, \label{wave} \\
\pi^{i j \text{TT}}_{\text{can}} &= \frac{1}{2} \dot{h}^{\text{TT}}_{i j}
	- \delta^{\text{TT} i j}_{k l}  \left( V^k_{(3)} \phi_{(2),l} - \pi^{kl}_{(5)\text{matter}} \right) \,, \label{pitt}
\end{align}
with the definition:
\begin{equation}\label{Bdef}
B_{(6)ij} = \hat{B}_{(6)ij} + \dot{\pi}^{ij}_{(5)\text{matter}} \,.
\end{equation}

One can get alternative expressions for $B_{(4)ij}$ and $B_{(6)ij}$ in terms of
$\mathcal{T}_{i j} = \sqrt{\gamma} T_{i j}$ by comparing the wave equation for
$h^{\text{TT}}_{i j}$ with the Einstein equations (see \cite{Steinhoff:Wang:2009}),
e.g.,
\begin{equation}\label{B4}
B_{(4)ij} = - 8\pi G \mathcal{T}_{(4) ij}
	- \frac{1}{8} \phi_{(2) , i} \phi_{(2) , j} \,.
\end{equation}
This should agree with Eq.\ (\ref{B4ham}) after the TT-projection.

\subsection{Near-zone expansion}
At the considered order, aspect like tail effects play no role (see
e.g.\ \cite{Blanchet:Damour:1988}). We may therefore solve the wave equation for
$h^{\text{TT}}_{i j}$ by an order-by-order evaluation of the retarded solution.
Further, the field solution is only needed in the near-zone.

In order to discuss the near-zone expansion, we write the wave equation for
$h^{\text{TT}}_{i j}$ schematically as:
\begin{equation}
\Box h^{\text{TT}}_{i j} = - 8\pi G \delta^{\text{TT} i j}_{k l} S_{kl} \,.
\end{equation}
The near-zone expansion of the retarded solution to this equation corresponds
to a series in $c^{-1}$ entering through the retarded time
$t_{\text{ret}} = t - c^{-1} | \vct{x} - \vct{x}^{\prime} |$, reading
\begin{equation}
\begin{split}
h^{\text{TT}}_{i j} &= - 8\pi G \delta^{\text{TT} i j}_{k l} \bigg[
	L_0 S_{kl} - L_1 \dot{S}_{kl} \nlq
	+ L_2 \ddot{S}_{kl}
	- L_3 \dddot{S}_{kl} + \dots \bigg] \,, \label{NZexpansion}
\end{split}
\end{equation}
where the TT-projector was pulled in front of the retarded solution and the
integral operator $L_n$ is defined by
\begin{equation}\label{Lop}
(L_n f)(\vct{x}, t) = - \frac{1}{4 \pi n!} \intxp \, | \vct{x} - \vct{x}^{\prime} |^{n-1} f(\vct{x}^{\prime}, t) \,.
\end{equation}
Notice that $L_{2n} = \Delta^{-1-n}$ for $n \in \mathbb{N}$, in particular
$L_0 = \Delta^{-1}$.

Using the PN-expanded source of the wave equation from Eq.\ (\ref{wave}) one may
arrange the near-zone expansion by PN orders as:
\begin{equation}\label{httexp}
h^{\text{TT}}_{i j} = h^{\text{TT}}_{(4) i j} + h^{\text{TT}}_{(5) i j} + h^{\text{TT}}_{(6) i j} + h^{\text{TT}}_{(7) i j} + \dots \,.
\end{equation}
It is important that only a finite number of terms from the near-zone expansion
[Eq.\ (\ref{NZexpansion})] contribute to a specific PN order due to the increasing
number of time derivatives therein. Therefore, one obtains
\begin{align}
h^{\text{TT}}_{(4) i j} &= 2 \delta^{\text{TT} i j}_{k l} \Delta^{-1} B_{(4)kl} \,, \label{htt4} \\
h^{\text{TT}}_{(5) i j} &= \dot{\chi}_{(4)ij} \,, \label{htt5} \\
h^{\text{TT}}_{(7) i j} &= \dot{\Pi}_{1 ij} + \dot{\Pi}_{2 ij}
	+ \ddot{\Pi}_{3 ij} + \dddot{\Pi}_{4 ij} + Q_{ij} \,, \label{htt7}
\end{align}
where
\begin{align}
\chi_{(4)ij} &= - 2 \delta^{\text{TT} i j}_{k l} L_1 B_{(4)kl} \,, \label{chi4old} \\
\Pi_{1 ij} &= - 2 \delta^{\text{TT} i j}_{k l} L_1 B_{(6)kl} \,, \\
\Pi_{2 ij} &= 8\pi G \delta^{\text{TT} i j}_{k l} L_1 \left( h^{\text{TT}}_{(4) kl} \mathcal{H}^{\rm matter}_{(2)} \right) , \\
\Pi_{3 ij} &= 2 \delta^{\text{TT} i j}_{k l} L_1 \left( V^k_{(3)} \phi_{(2),l} \right) ,  \label{PI3old} \\
\Pi_{4 ij} &= - 2 \delta^{\text{TT} i j}_{k l} L_3 B_{(4)kl} \,, \\
Q_{ij} &= - 8\pi G \delta^{\text{TT} i j}_{k l} \Delta^{-1} \left( h^{\text{TT}}_{(5) kl} \mathcal{H}^{\rm matter}_{(2)} \right) .
\end{align}
Notice that the application of $L_1$ to a total divergence like
$( \phi_{(2)} h^{\text{TT}}_{kl,m} )_{,m}$ leads to a vanishing result.
It will become apparent in the next section that $h^{\text{TT}}_{(6) i j}$ is
not needed in the present paper (but it contributes to the conservative 3PN
Hamiltonian). The definitions $P_{1ij} = \dot{\Pi}_{1ij}$,
$P_{2ij} = \dot{\Pi}_{2ij}$, $P_{3ij} = \ddot{\Pi}_{3ij}$, and
$R_{ij} = \dddot{\Pi}_{4 ij}$ were used in \cite{Jaranowski:Schafer:1997}.

An application of the operator $L_1$ obviously leads to a field depending on
time only (i.e., not depending on $\vct{x}$). This allows an easy calculation of
the (regularized) TT-projections in Eqs.\ (\ref{chi4old}) -- (\ref{PI3old}) by means of
the formula
\begin{equation}\label{TTtoSTF}
\delta^{\text{TT} i j}_{k l} A_{kl}(t) = \frac{2}{5} A_{ij}^{\text{STF}}(t) \,,
\end{equation}
valid for an arbitrary $\vct{x}$-independent function $A_{kl}(t)$ (see
\cite{Jaranowski:Schafer:1997}). Here STF denotes the symmetric trace-free part,
\begin{equation}
A_{ij}^{\text{STF}} = \frac{1}{2} ( A_{ij} + A_{ji} ) - \frac{1}{3} \delta_{ij} A_{kk} \,.
\end{equation}
Further, $h^{\text{TT}}_{(5) i j}$ is a function of time only,
$h^{\text{TT}}_{(5) i j , k} = 0$. As a consequence of these simplifications,
we finally have
\begin{align}
\chi_{(4)ij} &= \frac{1}{5 \pi} \intx \, B_{(4)ij}^{\text{STF}} \,, \label{chi4} \\
\Pi_{1 ij} &= \frac{1}{5 \pi} \intx \, B_{(6)ij}^{\text{STF}} \,, \label{PI1} \\
\Pi_{2 ij} &= - \frac{4 G}{5} \intx \, h^{\text{TT}}_{(4) ij} \mathcal{H}^{\rm matter}_{(2)} \,, \label{PI2} \\
\Pi_{3 ij} &= - \frac{1}{5 \pi} \intx \left( V^i_{(3)} \phi_{(2),j} \right)^{\text{STF}} , \label{PI3} \\
\Pi_{4 ij} &= \frac{1}{12 \pi} \delta^{\text{TT} i j}_{k l} \intxp \, | \vct{x} - \vct{x}^{\prime} |^2 B_{(4)kl} (\vct{x}^{\prime}, t) \,, \label{PI4} \\
Q_{ij} &= \frac{1}{2} h^{\text{TT}}_{(5) kl} \delta^{\text{TT} i j}_{k l} \phi_{(2)} \,, \label{Qij}
\end{align}
where the PN-expanded Hamilton constraint in the form
$\Delta \phi_{(2)} = - 16\pi G \mathcal{H}^{\rm matter}_{(2)}$ was used to
arrive at the last equation.

\subsection{Radiation-reaction Hamiltonians}
The dissipation through emission of gravitational radiation enters the PN-expansion
via $h^{\text{TT}}_{(5) i j}$ and $h^{\text{TT}}_{(7) i j}$, which are
antisymmetric under time reversal. The parts of the Hamiltonian linear in
$h^{\text{TT}}_{(5) i j}$ or $h^{\text{TT}}_{(7) i j}$ thus give the
radiation-reaction Hamiltonians at the considered order. Notice that
$H^{\text{field}}$ does not contribute to the matter equations of motion, so we
only need to consider $H^{\text{int}}$. The radiation-reaction Hamiltonians are thus
given by:
\begin{align}\label{Hint25}
H^{\text{int}}_{\text{2.5PN}} &= \frac{1}{16\pi G} \int \dd^3 x \, B_{(4)ij} h^{\text{TT}}_{(5) i j} \,, \\
H^{\text{int}}_{\text{3.5PN}} &= \frac{1}{16\pi G} \int \dd^3 x \, \bigg[
	B_{(4)ij} h^{\text{TT}}_{(7) i j}
	+ V^i_{(3)} \phi_{(2),j} \dot{h}^{\text{TT}}_{(5) i j} \nnlq
	+ \left( B_{(6)ij} - 4\pi G \mathcal{H}^{\rm matter}_{(2)} h^{\text{TT}}_{(4) i j} \right) h^{\text{TT}}_{(5) i j} \bigg] \nnl
	- \frac{1}{16\pi G} \frac{d}{d t} \int \dd^3 x \, h^{\text{TT}}_{(5) i j} \pi^{i j}_{(5)\text{matter}} \,, \label{Hint35}
\end{align}
where we used $h^{\text{TT}}_{(5) i j , k} = 0$, with Eqs.\ (\ref{Bdef}) and (\ref{pitt}).
Equation (\ref{pitt}) reads explicitly:
\begin{equation}
\pi^{i j \text{TT}}_{(6) \text{can}} = \frac{1}{2} \dot{h}^{\text{TT}}_{(5) i j} \,.
\end{equation}
The last term in Eq.\ (\ref{Hint35}) corresponds to a canonical transformation and
could be dropped, but we keep it for now.

One has to be aware of a subtlety here. The matter variables entering the
Hamiltonian via the solution for $h^{\text{TT}}_{i j}$ play a special role as
they may not be treated as dynamical (i.e., phase space) variables. Otherwise,
the matter equations of motion resulting from the Hamiltonian would in general
be wrong (at the conservative level one can use a Routhian to avoid this
problem, see \cite{Jaranowski:Schafer:1998}). Instead these nondynamical matter
variables entering through $h^{\text{TT}}_{i j}$ are treated as \emph{functions}
depending explicitly on time only. This introduces an explicit time-dependence
into the radiation-reaction Hamiltonians, which is a very natural description
of a dissipative system via canonical methods.

In order to distinguish the nondynamical matter variables from the dynamical
ones, we attach a prime to their object label as in, e.g., $\vct{P}_{1^\prime}$ or
$\vct{P}_{a^\prime}$, and also talk of primed and unprimed variables for short.
Further, we introduce an explicit time derivative $\dtex$, which only acts on the
primed variables (The partial and ordinary time derivatives act on both primed
and unprimed variables here). A superscript $\primevarsc$ is attached to a field
to denote that its solution should be expressed in terms of the primed
variables. This denotes an exchange of \emph{all} object labels by labels with
a prime, not just of label $a$. Thus $h^{\text{TT}}_{(5) i j}$ and
$h^{\text{TT}}_{(7) i j}$  in Eqs.\ (\ref{Hint25}) and (\ref{Hint35}) should better be
denoted by $h^{\text{TT}}_{(5) i j}\primevar{{}}$ and
$h^{\text{TT}}_{(7) i j}\primevar{{}}$ from now on. After the equations of
motion have been obtained from the Hamiltonian, one may identify primed and
unprimed variables (e.g., the objects $1$ and $1^{\prime}$), which in general
requires another application of regularization techniques.

The formulas for the radiation-reaction Hamiltonians Eqs.\ (\ref{Hint25}) and 
(\ref{Hint35}) can be simplified further. First, Eq.\ (\ref{Hint25}) may be
written as:
\begin{equation}\label{Hint25simp}
H^{\text{int}}_{\text{2.5PN}} = \frac{1}{16\pi G} h^{\text{TT}}_{(5) i j}\primevar{{}} \int \dd^3 x \, B_{(4)ij}^{\text{STF}} \,,
\end{equation}
where the $\vct{x}$-independent $h^{\text{TT}}_{(5) i j}$ was pulled in front of
the integral and $B_{(4)ij}$ is contracted with the symmetric
trace-free $h^{\text{TT}}_{(5) i j}$. As explained previously,
$h^{\text{TT}}_{(5) i j}$ must be replaced by
$h^{\text{TT}}_{(5) i j}\primevar{{}}$. The remaining integral in
Eq.\ (\ref{Hint25simp}) is identical up to a prefactor to the definition of
$\chi_{(4)ij}$, cf.\ (\ref{chi4}). Finally we obtain, inserting Eq.\ (\ref{htt5}),
\begin{equation}\label{Hrr25}
H^{\text{int}}_{\text{2.5PN}} = \frac{5}{16 G} \primevar{\dot{\chi}_{(4)ij}} \chi_{(4)ij} \,,
\end{equation}
which is a well-known result (see \cite{Jaranowski:Schafer:1997} and references
therein). The problem was reduced to the calculation of $\chi_{(4)ij}$ via
(\ref{chi4}). Remember that $\primevar{\dot{\chi}_{(4)ij}}$ in this Hamiltonian
is explicitly time-dependent.

We proceed with a simplification of the individual parts of Eq.\ (\ref{Hint35}).
Analogous to the simplification of $H^{\text{int}}_{\text{2.5PN}}$ given in the
last paragraph we have
\begin{gather}
\frac{1}{16\pi G} \int \dd^3 x \, h^{\text{TT}}_{(5) i j}\primevar{{}} B_{(6)ij} =
	\frac{5}{16 G} \primevar{\dot{\chi}_{(4)ij}} \Pi_{1 ij} \,, \label{Hint35_1} \\
\frac{1}{16\pi G} \int \dd^3 x \, \dot{h}^{\text{TT}}_{(5) i j}\primevar{{}} V^i_{(3)} \phi_{(2),j} =
	- \frac{5}{16 G} \primevar{\ddot{\chi}_{(4)ij}} \Pi_{3 ij} \,, \label{Hint35_2}
\end{gather}
where Eqs.\ (\ref{PI1}) and (\ref{PI3}) were used. We may further write
\begin{equation}\label{Hint35_3}
- \frac{1}{4} \int \dd^3 x \, h^{\text{TT}}_{(5) i j}\primevar{{}}
	h^{\text{TT}}_{(4) i j}\primevar{{}} \mathcal{H}^{\rm matter}_{(2)} =
	\frac{5}{16 G} \primevar{\dot{\chi}_{(4)ij}} \widetilde{\Pi}_{2 ij} ,
\end{equation}
with the definition
\begin{equation}\label{PI2Tdef}
\widetilde{\Pi}_{2 ij} = - \frac{4 G}{5} \int \dd^3 x \, h^{\text{TT}}_{(4) i j}\primevar{{}} \mathcal{H}^{\rm matter}_{(2)} \,.
\end{equation}
The notation $\widetilde{\Pi}_{2 ij}$ was chosen because of the similarity to
$\Pi_{2 ij}$, cf.\ Equation (\ref{PI2}). If the self-interaction contributions to the
integral in (\ref{PI2}) vanish, then $\widetilde{\Pi}_{2 ij}$ can be obtained
from $\Pi_{2 ij}$ by a relabeling of objects only. For the spin-dependent part of $\widetilde{\Pi}_{2 ij}$, this will turn out to
be possible.
The integral over $B_{(4)ij} h^{\text{TT}}_{(7) i j}$ in Eq.\ (\ref{Hint35}) splits
into the following five parts, cf.\ Equation (\ref{htt7}),
\begin{align}
\frac{1}{16\pi G} \int \dd^3 x \, \primevar{\dot{\Pi}_{1 ij}} B_{(4)ij} &=
	\frac{5}{16 G} \primevar{\dot{\Pi}_{1 ij}} \chi_{(4)ij} \,, \label{Hint35_4} \\
\frac{1}{16\pi G} \int \dd^3 x \, \primevar{\dot{\Pi}_{2 ij}} B_{(4)ij} &=
	\frac{5}{16 G} \primevar{\dot{\Pi}_{2 ij}} \chi_{(4)ij} \,, \label{Hint35_5} \\
\frac{1}{16\pi G} \int \dd^3 x \, \primevar{\ddot{\Pi}_{3 ij}} B_{(4)ij} &=
	\frac{5}{16 G} \primevar{\ddot{\Pi}_{3 ij}} \chi_{(4)ij} \,, \label{Hint35_6} \\
\frac{1}{16\pi G} \int \dd^3 x \, \primevar{\dddot{\Pi}_{4 ij}} B_{(4)ij} &=
	(\dtex)^3 ( R^{\prime} + R^{\prime\prime} ) \,, \label{Hint35_7} \\
\frac{1}{16\pi G} \int \dd^3 x \, \primevar{Q_{ij}} B_{(4)ij} &=
	\primevar{\dot{\chi}_{(4)ij}} ( Q^{\prime}_{ij} + Q^{\prime\prime}_{ij} ) \label{Hint35_8} \,.
\end{align}
Notice that here $\Pi_{1 ij}$, $\Pi_{2 ij}$, and $\Pi_{3 ij}$ are
independent of $\vct{x}$. The relations Eqs.\ (\ref{Qij}) and (\ref{htt5}) were used in
the last integral. The last two integrals were each split into two parts using
Eq.\ (\ref{B4}) and the following definitions:
\begin{align}
R^{\prime} &= - \frac{1}{2} \int \dd^3 x \,
	\mathcal{T}_{(4) ij} \primevar{\Pi_{4 ij}} \,, \label{Rpdef} \\
R^{\prime\prime} &= -\frac{1}{128\pi G} \int \dd^3 x \,
	\phi_{(2) , i} \phi_{(2) , j} \primevar{\Pi_{4 ij}} \,, \label{Rppdef} \\
Q^{\prime}_{ij} &= - \frac{1}{4} \int \dd^3 x \,
	\mathcal{T}_{(4) kl} \delta^{\text{TT} kl}_{ij} \primevar{\phi_{(2)}} \,, \label{Qpdef} \\
Q^{\prime\prime}_{ij} &= -\frac{1}{256\pi G} \int \dd^3 x \,
	\phi_{(2) , k} \phi_{(2) , l} \delta^{\text{TT} kl}_{ij} \primevar{\phi_{(2)}} \,. \label{Qppdef}
\end{align}
The fact that the explicit time derivative $\dtex$ only acts on primed variables
was used in (\ref{Hint35_7}) to pull it in front of the whole expression.
Finally, it holds that 
\begin{equation}\label{Hint35_9}
- \frac{1}{16\pi G} \int \dd^3 x \, h^{\text{TT}}_{(5) i j}\primevar{{}} \pi^{i j}_{(5)\text{matter}} =
	- \primevar{\dot{\chi}_{(4)ij}} O_{ij} \,,
\end{equation}
with the definition
\begin{equation}\label{defO}
O_{ij} = \frac{1}{16\pi G}  \int \dd^3 x \, \pi^{i j}_{(5)\text{matter}} \,.
\end{equation}

Summing up the contributions from Eqs.\ (\ref{Hint35_1}) -- (\ref{Hint35_3}),
(\ref{Hint35_4}) -- (\ref{Hint35_8}), and the total time derivative of
Eq.\ (\ref{Hint35_9}), one gets:
\begin{align}
\begin{split}
H^{\text{int}}_{\text{3.5PN}} &= \frac{5}{16 G} \Big[
		\chi_{(4)ij} ( \primevar{\dot{\Pi}_{1 ij}}
			+ \primevar{\dot{\Pi}_{2 ij}}
			+ \primevar{\ddot{\Pi}_{3 ij}} ) \nlq
		+ \primevar{\dot{\chi}_{(4)ij}} ( \Pi_{1 ij} + \widetilde{\Pi}_{2 ij} )
		- \primevar{\ddot{\chi}_{(4)ij}} \Pi_{3 ij}
	\Big] \nl
	+ \primevar{\dot{\chi}_{(4)ij}} ( Q^{\prime}_{ij} + Q^{\prime\prime}_{ij} )
	+ (\dtex)^3 ( R^{\prime} + R^{\prime\prime} ) \nl
	- \frac{d}{d t} \Big[ \primevar{\dot{\chi}_{(4)ij}} O_{ij} \Big] \,. \label{Hrr35}
\end{split}
\end{align}
This agrees with \cite{Jaranowski:Schafer:1997} (with misprints corrected
in \cite{Konigsdorffer:Faye:Schafer:2003}). It should be noted that no time
derivatives are present in Eq.\ (\ref{Hint}), so all time derivatives in
Eqs.\ (\ref{Hrr25}) and (\ref{Hrr35}) are introduced by above insertions. Indeed, all
these time derivatives should be understood as abbreviations and be performed
before the equations of motions are derived from the Hamiltonians. However, for
time derivatives of primed variables it is irrelevant at which stage they are
eliminated (These are actually all time derivatives except the one acting on
$O_{ij}$). One should be aware that an insertion of equations of motion leads to
a recombination of PN orders, e.g., inserting the 1PN conservative part of the equations
of motion leads to 3.5PN contributions from $H^{\text{int}}_{\text{2.5PN}}$,
cf.\ Equation (\ref{Hrr25}). Further, one should notice that $\widetilde{\Pi}_{2 ij}$, $R^{\prime}$,
$R^{\prime\prime}$, $Q^{\prime}_{ij}$, and $Q^{\prime\prime}_{ij}$ depend on
both primed and unprimed variables by virtue of their definitions.

\section{Calculation of the Hamiltonians\label{sec:calculation}}
Up to formal 3.5PN order, the interaction Hamiltonian is given by Eqs.\ (\ref{Hrr25})
and (\ref{Hrr35}). The quantities entering these expressions must be calculated
by solving the integrals appearing in their definitions [see Eqs.\ (\ref{htt4}),
(\ref{chi4}--\ref{PI4}), (\ref{PI2Tdef}), (\ref{Rpdef}--\ref{Qppdef}], and
(\ref{defO})).
The leading-order source terms in the pole-dipole case entering these integrals
read
\begin{align}
\mathcal{H}^{\rm matter}_{(2)} &= \sum_a m_a \delta_a \,, \label{H2} \\
\begin{split}\label{Tij}
\mathcal{T}_{(4) i j} &= \sum_a \frac{1}{m_a} \bigg[ P_{a i} P_{a j} \delta_a
	+ P_{a ( i} S_{a (j))(k)} \partial_k \delta_a \bigg] \,,
\end{split}\\
B_{(4)ij} &= - 8\pi G \mathcal{T}_{(4) ij}
	- \frac{1}{8} \phi_{(2) , i} \phi_{(2) , j} \,, \label{B4ij}
\end{align}
see \cite{Steinhoff:Wang:2009} for more details. Here, $m_a$ ($a=1,2 \dots$) are the masses and
$\delta_a = \delta(\vct{x}-\hat{\vct{z}}_a)$. $\phi_{(2)}$ is proportional to
the Newtonian potential of point-masses, namely:
\begin{equation}\label{Npotential}
\phi_{(2)} = - 16\pi G \Delta^{-1} \mathcal{H}^{\rm matter}_{(2)} = 4 G \sum_a \frac{m_a}{r_a} \,,
\end{equation}
where $r_a = | \vct{x}-\hat{\vct{z}}_a |$. Notice that $\phi_{(2)}$ is
independent of the spins. The expression for $B_{(6)ij}$ was derived in
\cite{Steinhoff:Wang:2009}:
\begin{widetext}
\begin{align}
\begin{split}
B_{(6)ij} &= 16\pi G \sum_a
	\bigg[ \frac{{\bf P}^2_a}{4m_a^3} P_{a i} P_{a j} \delta_a
	+ \frac{5}{8m_a} P_{a i} P_{a j} \phi_{(2)} \delta_a
	+ \frac{{\bf P}^2_a}{4m_a^3} P_{a i} S_{a (j) (k)} \delta_{a,k}
	- \frac{1}{4m_a^3} P_{a l} P_{a j} P_{a k} S_{a (l) (i)} \delta_{a,k} \nlq
	+ \frac{5}{8m_a} P_{a i} S_{a (j) (k)} \left( \phi_{(2)} \delta_a \right)_{,k}
	+ \frac{1}{2m_a} P_{a i} S_{a (k) (j)} \phi_{(2),k} \delta_a
	- \frac{1}{8m_a} P_{a k} S_{a (k) (i)} \phi_{(2),j} \delta_a \nlq
	+ \frac{1}{2} S_{a (k) (i)} \left( V^j_{(3),k} + V^k_{(3),j} \right) \delta_a \bigg] \nl
	+ \frac{1}{2} \phi_{1(4)} \phi_{(2) , i j}
	+ \frac{3}{8} \phi_{2(4)} \phi_{(2) , i j}
	+ \frac{5}{64} \phi_{(2)} \phi_{(2) , i} \phi_{(2) , j}
	+ 2 \tilde{\pi}^{j k}_{(3)} \left( \tilde{\pi}^k_{(3),i} - \tilde{\pi}^i_{(3),k} \right)
	+ 2 \tilde{\pi}^{i j}_{(3),k} V^k_{(3)}
	+ \frac{1}{2} \tilde{\pi}^{i j}_{(3)} \tilde{\pi}^k_{(3),k} \,.
\label{B6ij}
\end{split}
\end{align}
\end{widetext}
The field quantities entering Eq.\ (\ref{B6ij}) are equal to:
\begin{subequations}
\begin{align}
\phi_{1(4)} &= 2 G \sum_a \bigg[ \frac{\vct{P}_a^2}{m_a r_a}
	+ \frac{P_{a i} S_{a (i) (j)}}{m_a} \left( \frac{1}{r_a}
        \right)_{,j} \bigg] \,, \\
\phi_{2(4)} &= - 2 G^2 \sum_a \sum_{b\neq a} \frac{m_a m_b}{r_{ab} r_a}\,, \\
\tilde{\pi}^i_{(3)} &= G \sum_a \bigg[ 2 \frac{P_{a i}}{r_a}
	+ S_{a (i) (j)} \left( \frac{1}{r_a} \right)_{,j} \bigg] \,, \\ 
V^i_{(3)} &= G \sum_a \bigg[ 2 \frac{P_{a i}}{r_a} - \frac{1}{4} P_{a j} r_{a,ij}
	+ S_{a (i) (j)} \left( \frac{1}{r_a} \right)_{,j} \bigg] \,, \\
\tilde{\pi}^{ij}_{(3)} &= G \sum_a \bigg[
	2 P_{a i} \left( \frac{1}{r_a} \right)_{,j}
	+ 2 P_{a j} \left( \frac{1}{r_a} \right)_{,i} \nonumber \nl
	- \delta_{ij} P_{a k} \left( \frac{1}{r_a} \right)_{,k}
	- \frac{1}{2} P_{a k} r_{a,ijk} \nl
        - S_{a (k) (i)} \left( \frac{1}{r_a} \right)_{, k j}
	- S_{a (k) (j)} \left( \frac{1}{r_a} \right)_{, k i} \bigg] \,, \nonumber
\end{align}
\end{subequations}
where $r_{ab} = | \hat{\vct{z}}_a-\hat{\vct{z}}_b |$.
Notice that for nonspinning systems the result in \cite{Jaranowski:Schafer:1997}
is reproduced. Further notice that $\tilde{\pi}^i_{(3),i}$ does not depend on
spin. Finally, the spin correction to the field momentum is given by:
\begin{equation}
\pi^{i j}_{(5)\text{matter}} = - \sum_a \frac{4 \pi G}{m_a^2} P_{a k} P_{a (i} S_{a (j)) (k)} \delta_a \,,
\end{equation}
to the required order.

\subsection{Spin-dependent part of \texorpdfstring{$\htt_{ij}$}{hTT}\label{sec:hTTS}}
The explicit solutions for the point-mass, i.e., spin-independent, contributions
to $\htt_{ij}$ can be found in \cite{Schafer:1985, Jaranowski:Schafer:1997,
Konigsdorffer:Faye:Schafer:2003, Jaranowski:Schafer:1998} (but notice that
\cite{Jaranowski:Schafer:1997} contains some misprints). The spin part of
$\htt_{(4)ij}$, arising from the spin-dependent source terms in Eq.\ (\ref{Tij}) via
Eqs.\ (\ref{B4ij}) and (\ref{htt4}), has been computed in
\cite{Steinhoff:Schafer:Hergt:2008} and reads:
\begin{align}
\htts_{(4) i j} &=
		G \sum_a \frac{P_{a n} S_{a (k) (l)}}{m_a} \bigg[
			( 4 \delta_{k ( i} \delta_{j ) n} \partial_l
			- 2 \delta_{i j} \delta_{k n} \partial_l ) \frac{1}{r_a} \nnl
		+ ( \delta_{k n} \partial_i \partial_j \partial_l
			- 2 \delta_{k ( i} \partial_{j )} \partial_n \partial_l ) r_a
		\bigg]\,,
\end{align}
where we use the superscript ``spin'' to denote the spin-dependent part of a quantity
from now on.  In order to obtain the spin contributions to the
radiation-reaction Hamiltonian up to formal 3.5PN order, we also need to compute
the spin part of  $\htt_{(5)ij}$ and $\htt_{(7)ij}$. $\htts_{(5)ij}$ is easy to
compute. From Eqs.\ (\ref{chi4}), (\ref{B4ij}), and (\ref{Tij}), we have
\begin{equation}
\chi_{(4)ij}^{\text{spin}} 
		= - \frac{8 G}{5} \sum_a \! \left[ \frac{P_{ai}S_{a(j)(k)}}{m_a}
                    \intx \, \partial_k \delta_{a} \right]^{\text{STF}}
                \! = 0 \,, \label{chi4result}
\end{equation}
and thus also $\htts_{(5)ij} = 0$ [see Eq.\ (\ref{htt5})]. There is no spin
contribution to the 2.5PN $\htt_{ij}$, which is the reason why the leading-order source
terms (\ref{Tij}) are not sufficient to derive the leading-order
radiation-reaction Hamiltonian. $\htts_{(6)ij}$ would be more difficult to derive, 
but it is not needed in our calculation of the leading-order radiation-reaction
Hamiltonian with spins, so we will not discuss it in the present paper.

Analogous to Eq.\ (\ref{htt7}), we decompose the solution for $\htts_{(7)ij}$ into
several parts,
\begin{equation}
\htts_{(7)ij} = \dot{\Pi}^{\rm spin}_{1 ij} + \dot{\Pi}^{\rm spin}_{2 ij}
	+ \ddot{\Pi}^{\rm spin}_{3 ij} + \dddot{\Pi}_{4 ij}^{\rm spin} \,,
\end{equation}
where the following definitions are used:
\begin{align}
\Pi_{1 ij}^{\text{spin}} &= \frac{1}{5 \pi} \intx \, B_{(6)ij}^{\text{STF spin}} \,, \\
\Pi_{2 ij}^{\text{spin}} &= - \frac{4 G}{5} \intx \, h^{\text{TT spin}}_{(4) ij} \, \mathcal{H}^{\rm matter}_{(2)} \,, \label{PI2spin} \\
\Pi_{3 ij}^{\text{spin}} &= - \frac{1}{5 \pi} \intx \left( V^{i \text{ spin}}_{(3)} \phi_{(2),j} \right)^{\text{STF}} , \\
\Pi_{4 ij}^{\text{spin}} &= - \frac{2 G}{3} \delta^{\text{TT} i j}_{k l} \intxp \, | \vct{x} - \vct{x}^{\prime} |^2 \mathcal{T}_{(4) kl}^{\text{spin}} (\vct{x}^{\prime}, t) \,,
\end{align}
and obviously $Q_{ij}^{\rm spin}=0$, cf.\ Equations (\ref{PI1}) -- (\ref{Qij}) and
(\ref{B4ij}). These integrals yield the results:
\begin{widetext}
    \begin{align}
    \Pi^{\rm spin}_{1ij}
       &= \frac{4 G^2}{5} \sum_a \sum_{b \ne a} \bigg\{
    \frac{1}{r_{ab}^2} \Big[
       3 {(  {\mathbf n}_{ab} \cdot {\mathbf P}_b ) }  n_{ab}^{k}
		(  n_{ab}^{j}  S_a{}_{(i)}{}_{(k)} +  n_{ab}^{i}  S_a{}_{(j)}{}_{(k)} )
       - 3 P_b {}_{k} ( n_{ab}^{j} S_a {}_{(i)}{}_{(k)} + n_{ab}^{i}  S_a{}_{(j)}{}_{(k)} ) \nnl
       - 3 n_{ab}^{k} ( P_b {}_{j} S_a {}_{(i)}{}_{(k)} + P_b {}_{i}  S_a {}_{(j)}{}_{(k)} )
       + 4 ( 3 n_{ab}^{i}  n_{ab}^{j} - \delta{}_{i}{}_{j} ) n_{ab} ^{k} P_b {}_{l}  S_a{}_{(k)}{}_{(l)}
         \Big]
    + \frac{m_b}{ m_a }   \frac{1}{r_{ab}^2} \Big[
       P_a {}_{k} (  n_{ab}^{j}  S_a {}_{(i)}{}_{(k)} +  n_{ab}^{i}  S_a {}_{(j)}{}_{(k)} ) \nnl
       + ( 4 \delta{}_{i}{}_{j} - 6 n_{ab}^{i}  n_{ab}^{j} ) { n_{ab}^{k}  P_a {}_{l}  S_a {}_{(k)}{}_{(l)} }
       + 4  n_{ab}^{k} (  P_a {}_{j}  S_a{}_{(i)}{}_{(k)} +  P_a {}_{i}  S_a {}_{(j)}{}_{(k)} )
     \Big]
   - \frac{S_a{}_{(k)}{}_{(l)}}{r_{ab}^3} \Big[
       (  3  n_{ab}^{i}  n_{ab}^{j} - \delta{}_{i}{}_{j} ) {   S_b {}_{(k)}{}_{(l)} } \nnl
       + 3n_{ab}^k ( n_{ab}^j S_{b}{}_{(i)(l)} + n_{ab}^{i} S_{b}{}_{(j)(l)} )
       + 3 (\delta{}_{i}{}_{j} - 5 n_{ab}^{i}  n_{ab}^{j} ) { n_{ab}^{k}  n_{ab}^{n}
           S_b {}_{(n)}{}_{(l)} }
   \Big]
 \bigg\} , \label{p1sresult} \\
  \begin{split}
       \Pi^{\rm spin}_{2ij}
       &=-\frac{4 G^2}{5} \sum_a \sum_{b \ne a} \frac{m_b}{ m_a }
         \frac{1}{r_{ab}^2} \Big[
	 - 2 P_{ak} ( n_{ab}^{i}   S_{a(j)(k)} + n_{ab} ^{j}   S_{a (i)(k)} )
         + n_{ab} ^{k} ( {P_{ai}}  {S_{a(j)(k)}} + {P_{aj}}  {S_{a(i)(k)}} )
 \nl
         + 3 {(  {\mathbf n}_{ab} \cdot {\mathbf P}_a ) } n_{ab} ^{k} ( n_{ab} ^{i}  S_{a(j)(k)} +
           n_{ab}^j S_{a(i)(k)} )
         + ( \delta_{ij} + 3 n_{ab}^i n_{ab}^j ) { n_{ab} ^{k}  P_{a l}  S_{a (k)(l)} }
            \Big]  \,,   \label{p2sresult}
  \end{split}\\
       \Pi^{\rm spin}_{3ij}
       &= \frac{4 G^2}{5}
       \sum_a \sum_{b \ne a} \frac{m_b}{{r_{ab}}}  
       n_{ab} ^{k} ( n_{ab} ^{j} {S_a}{}_{(i)}{}_{(k)} + n_{ab}^{i}
       {S_a}{}_{(j)}{}_{(k)}) \,,  \label{p3sresult} \\
\begin{split}
  \Pi^{\rm spin}_{4ij}
     &= 
     \frac{4 G}{15} \sum_a \frac{r_a}{ m_a }
     \Big[ P_a{}_{k} ( n_a^{j} S_a{}_{(i)}{}_{(k)} + n_a^{i} S_a{}_{(j)}{}_{(k)} )
       -2 n_a^{k} (  P_a {}_{j}  S_a {}_{(i)}{}_{(k)} +  P_a{}_{i}  S_a {}_{(j)}{}_{(k)}
            + \delta{}_{i}{}_{j} P_a {}_{l}  S_a{}_{(k)}{}_{(l)} )
      \Big] \,, \label{p4sresult}
\end{split}
\end{align}
\end{widetext}
where $\vct{n}_a = ( \vct{x} - \hat{\vct{z}}_a ) / r_a$ and
$\vct{n}_{ab} = ( \hat{\vct{z}}_a - \hat{\vct{z}}_b ) / r_{ab}$. Notice that it
holds
\begin{equation}
\Pi^{\rm spin}_{1ij} + \Pi^{\rm spin}_{2ij} + \dot{\Pi}^{\rm spin}_{3ij} = - \frac{4 G}{5} I_{ij}^{\text{spin}}\,,
\end{equation}
at the considered PN order, where $I_{ij}$ is a multipole moment of the far-zone
expansion of $\htt_{ij}$ and can be expressed as a double time derivative of a
very compact expression, see Eqs.\ (6.15) and (6.18) in \cite{Steinhoff:Wang:2009}.

\subsection{Derivation of spin contributions to 2.5PN and 3.5PN interaction Hamiltonians}
When taking into account the fact that Eq.\ (\ref{chi4result}) tells us that
$\chi_{(4)ij}^{\rm spin} = 0$, we immediately see that the formal
2.5PN order interaction Hamiltonian Eq.\ (\ref{Hrr25}),
\begin{equation}\label{eq:H25}
H^{\text{int}}_{\text{2.5PN}} = \frac{5}{16 G} \primevar{\dot{\chi}_{(4)ij}} \chi_{(4)ij} \,,
\end{equation}
has only the well-known point-mass contribution \cite{Schafer:1985,Schafer:1986}:
\begin{equation}
\begin{split}
\chi_{(4)ij} &= \frac{4 G}{15} \suma \bigg[ \frac{2}{m_a} ( \vct{P}_a^2 \delta_{ij} - 3 P_{ai} P_{aj} ) \nlq
	- G \sumb \frac{m_a m_b}{r_{ab}} ( \delta_{ij} - 3 n^i_{ab} n^j_{ab} ) \bigg] \,,
\end{split}
\end{equation}
but no \emph{direct} spin contribution. However, indirect spin-contributions
arise from Eq.\ (\ref{eq:H25}) via the time derivative therein and first appear at
the formal 3.5PN level [after taking into account the leading-order conservative
spin-orbit and spin(1)-spin(2) equations of motion \cite{Barker:OConnell:1975,
DEath:1975, Barker:OConnell:1979}, provided in this paper by Eqs.\ (\ref{eq:xdot}) and
(\ref{eq:pdot}) later on].

The spin part of the formal 3.5PN order interaction Hamiltonian Eq.\ (\ref{Hrr35})
can be written as:
\begin{align}
H^{\text{int spin}}_{\text{3.5PN}} &= \frac{5}{16 G} \Big[
		( \dot{\Pi}_{1 ij}^{\text{spin}}
			+ \dot{\Pi}_{2 ij}^{\text{spin}}
			+ \ddot{\Pi}_{3 ij}^{\text{spin}} )\primevar{{}} \chi_{(4)ij} \nonumber \nlq
		+ \primevar{\dot{\chi}_{(4)ij}} ( \Pi_{1 ij}^{\text{spin}} + \widetilde{\Pi}_{2 ij}^{\text{spin}} )
		- \primevar{\ddot{\chi}_{(4)ij}} \Pi_{3 ij}^{\text{spin}}
	\Big] \nonumber \nl
	+ \primevar{\dot{\chi}_{(4)ij}} Q^{\prime\text{spin}}_{ij}
	+ (\dtex)^3 ( R^{\prime\text{spin}} + R^{\prime\prime\text{spin}} ) \nonumber \nl
	- \frac{d}{d t} \Big[ \primevar{\dot{\chi}_{(4)ij}} O_{ij}^{\text{spin}} \Big] \,, \label{eq:H35final}
\end{align}
where we used $\chi_{(4)ij}^{\rm spin} = 0$ and
$Q^{\prime\prime\text{spin}}_{ij} = 0$. The latter is trivial from
Eq.\ (\ref{Qppdef}), as only the spin-independent potential $\phi_{(2)}$ appears
there. $\Pi^{\rm spin}_{1ij}$, $\Pi^{\rm spin}_{2ij}$, $\Pi^{\rm spin}_{3ij}$,
and $\Pi^{\rm spin}_{4ij}$ were already derived in Sec. \ref{sec:hTTS}. The
missing quantities $\widetilde{\Pi}_{2 ij}^{\text{spin}}$,
$Q^{\prime\text{spin}}_{ij}$, $R^{\prime\text{spin}}$,
$R^{\prime\prime\text{spin}}$, and $O_{ij}^{\text{spin}}$ can be obtained from:
\begin{align}
\widetilde{\Pi}_{2 ij}^{\text{spin}} &= - \frac{4 G}{5} \int \dd^3 x \, h^{\text{TT spin}}_{(4) i j}\primevar{{}} \mathcal{H}^{\rm matter}_{(2)} \,. \label{PI2Tspin} \\
Q^{\prime\text{spin}}_{ij} &= - \frac{1}{4} \int \dd^3 x \,
	\mathcal{T}_{(4) kl}^{\text{spin}} \delta^{\text{TT} kl}_{ij} \primevar{\phi_{(2)}} \,, \label{Qpspin} \\
R^{\prime\text{spin}} &= - \frac{1}{2} \int \dd^3 x \,
	( \mathcal{T}_{(4) ij} \primevar{\Pi_{4 ij}} )^{\text{spin}} \,,  \\
R^{\prime\prime\text{spin}} &= - \frac{1}{8} \frac{1}{16\pi G} \int \dd^3 x \,
	\phi_{(2) , i} \phi_{(2) , j} \Pi_{4 ij}^{\text{spin}}\primevar{{}} \,, \\
O_{ij}^{\text{spin}} &= \frac{1}{16\pi G} \int \dd^3 x \, \pi^{i j \text{ spin}}_{(5)\text{matter}} \,,
\end{align}
using Eqs.\ (\ref{PI2Tdef}), (\ref{Rpdef}) -- (\ref{Qppdef}), and (\ref{defO}). We also
split $R^{\prime\text{spin}}$ into three parts,
\begin{align}
R^{\prime\text{spin}} &= R^{\prime\text{spin}}_1 + R^{\prime\text{spin}}_2 + R^{\prime\text{spin}}_3 \,, \\
R^{\prime\text{spin}}_1 &= - \frac{1}{2} \int \dd^3 x \,
	\mathcal{T}_{(4) ij}^{\text{spin}} \Pi_{4 ij}^{\text{PM}}\primevar{{}} \,, \\
R^{\prime\text{spin}}_2 &= - \frac{1}{2} \int \dd^3 x \,
	\mathcal{T}_{(4) ij}^{\text{PM}} \Pi_{4 ij}^{\text{spin}}\primevar{{}} \,, \\
R^{\prime\text{spin}}_3 &= - \frac{1}{2} \int \dd^3 x \,
	\mathcal{T}_{(4) ij}^{\text{spin}} \Pi_{4 ij}^{\text{spin}}\primevar{{}} \,.
\end{align}
Here, PM denotes the point-mass parts of a function. The quantities entering above integrals
will be all given in the present paper, except for $\Pi_{4 ij}^{\text{PM}}$, which
can be read from Eq.\ (36) in \cite{Jaranowski:Schafer:1997} using
$R_{ij} = \partial_t^3 \Pi_{4 ij}$. The results of the above integrations read as follows:
\begin{widetext}
\begin{align}
\begin{split}
  {\widetilde \Pi}_{2ij}^{\rm spin}
  &= -\frac{4 G^2}{5} \sum_{a, a^{\prime}} \frac{m_a}{ \mapr }
         \frac{1}{\raap^2} \Big[
	 2  P_{a^\prime k} ( \naap^{i} S_{a^\prime (j)(k)}
         + \naap ^{j}   S_{a^\prime (i)(k)} )
         - \naap ^{k} ( {P_{a^\prime i}}  {S_{a^\prime (j)(k)}} + {P_{a^\prime j}}  {S_{a^\prime (i)(k)}} ) \nl
         - 3 {(  {\mathbf n}_{{a} a^\prime} \cdot {\mathbf P}_{a^\prime} ) }   \naap ^{k} ( \naap ^{i}  S_{a^\prime (j)(k)} +
           \naap^j S_{a^\prime (i)(k)} )
         - ( \delta_{ij} + 3 \naap^i \naap^j ) { \naap ^{k}  P_{a^\prime l}  S_{a^\prime (k)(l)} } \Big] \,, \label{pi2tresult}
\end{split}\\
  \begin{split}
     Q_{ij}^{\prime \rm spin}
     &= \frac{G}{4} \sum_{a, a^{\prime}} \frac{\mapr}{ m_a }
    \frac{1}{\raap^2}  
    \Big[
      2 P_a {}_{k} ({   \naap^{i}    S_a {}_{(j)}{}_{(k)}  }+{   \naap^{j}    S_a {}_{(i)}{}_{(k)}  } ) 
      - \naap^{k} ({  P_a {}_{i}
          S_a {}_{(j)}{}_{(k)}  }+{     P_a {}_{j}
          S_a {}_{(i)}{}_{(k)}  } )
\nl
      - 3  {(  {\mathbf n}_{a {a^\prime}} \cdot  {\mathbf P}_a )} \naap^{k} (
      {   \naap^{i}    S_a
          {}_{(j)}{}_{(k)}  }+
      {   \naap^{j}    S_a
          {}_{(i)}{}_{(k)}  } )
       - (\delta_{ij}+3\naap^{i} \naap^{j} )\naap^{k}  P_{a} {}_{l}  S_{a} {}_{(k)}{}_{(l)} \Big] \,, \label{Qpresult}
  \end{split}\\
R^{\prime\text{spin}}_1
      &=
      \frac{G}{15} \sum_{a, a^{\prime}} S_a {}_{(i)}{}_{(j)} \bigg( \frac{4 r_{a^{\prime} a}}{ m_{a^\prime} m_a }
       \Big[ 
       {  {\bf P}_{a^\prime}^2   }
        {  n_{{a^\prime} a}^{i}  P_a {}_{j}  }
      - ( {\bf n}_{{a^\prime} a} \cdot  {\bf P}_{a^\prime} )
           {  P_{a^\prime} {}_{i}  P_a {}_{j} }
        - 2 {(  {\bf P}_{a^\prime} \cdot
        {\bf P}_a  ) }  {  n_{{a^\prime} a}^{i}
        P_{a^\prime} {}_{j}  }  \Big] \nnl
      + \frac{G}{7} \sumbp \frac{m_{a^\prime}
        m_{b^\prime}}{ m_a }  \bigg\{
       17 { n_{a^{\prime} b^{\prime}}^{i}  P_a{}_{j}  }
      - \frac{2 r_{a^{\prime} a}}{r_{a^{\prime}
          b^{\prime} } }  \Big[
       17 {( {\bf n}_{a^{\prime} b^{\prime}} \cdot  {\bf P}_a  ) }  
        { n_{a^{\prime} b^{\prime}}^{i}  n_{{a^\prime} a}^{j}  } 
       + 7 { n_{{a^\prime}
          a} {}^{i}  P_a {}_{j}  }
      \Big] \label{RS1result} \nl
        + \frac{6r_{a^{\prime} a}^2}{r_{a^{\prime} b^{\prime}}^2} 
        \Big[ 
        { n_{a^{\prime} b^{\prime}}^{i}  P_a {}_{j} }
       + 2 {({\bf n}_{{a^\prime} a} \cdot  {\bf P}_a ) }  
        { n_{a^{\prime} b^{\prime}}^{i}  n_{{a^\prime} a}^{j} }
        \Big]
      + \frac{8 r_{a^{\prime} a}}{r_{a^{\prime} b^{\prime} }^3} \Big[
      r_{a^{\prime} a}^{2} { n_{{a^\prime} a}^{i}  P_a
        {}_{j}}  
      - r_{b^{\prime} a}^{2} { n_{{a^\prime} a}^{i}  P_a {}_{j}  }  
      \Big] 
 \bigg\} \bigg) \,, \nonumber \\
   \begin{split}
R^{\prime\text{spin}}_2
      &= 
     \frac{4 G}{15}  \sum_{a, a^{\prime}} \frac{r_{a a^{\prime}}}{ m_{a^\prime}
       m_a }  S_{a^{\prime}} {}_{(i)}{}_{(j)}
\Big[  {  {\bf P}_a^2  }  n_{a a^{\prime}}^{i}
       P_{a^{\prime}} {}_{j} -2{  ({\bf P}_{a^{\prime}}\cdot{\bf P}_a  })  n_{a a^{\prime}}^{i}
       P_{a} {}_{j}
     +{  ( \vct{n}_{a a^{\prime}} \cdot{\bf P}_{a}})P_{a^{\prime}} {}_{i}P_{a} {}_{j} \Big] \,,
   \end{split}\\
R^{\prime\text{spin}}_3
      &=
      \frac{4 G}{15} \suma \sum_{a^{\prime} \neq a}
      \frac{1}{ m_{a^\prime} m_a } S_a {}_{(i)}{}_{(j)} \bigg[
       \frac{3}{2} { P_{a^\prime} {}_{k}  P_a {}_{i}  S_{a^\prime}
            {}_{(k)}{}_{(j)} }
       - {(  {\bf P}_{a^\prime} \cdot
        {\bf P}_a ) }  { S_{a^\prime} {}_{(i)}{}_{(j)} }  
        - { P_{a^\prime} {}_{i}  P_a {}_{k}  S_{a^\prime}
          {}_{(k)}{}_{(j)} }  \bigg] \,, \\
   \begin{split}
R^{\prime\prime\text{spin}}
      &=
      \frac{2 G^2}{15} \sum_{a, a^{\prime}} \sumb \frac{m_a
       m_b}{ m_{a^\prime} }  \frac{r_{a^{\prime} a}}{r_{ab}} S_{a^\prime} {}_{(i)}{}_{(j)} \Big[
       { n_{{a^\prime} a} {}^{i}
       P_{a^\prime} {}_{j} }
     - 2 {(  {\bf n}_{ab} \cdot  {\bf P}_{a^\prime}  ) }
       { n_{{a^\prime} a}^{i}
       n_{ab}^{j} }  
     - { ( {\bf n}_{{a^\prime} a} \cdot
       {\bf n}_{ab} ) }  { n_{ab} {}^{i}  P_{a^\prime}
       {}_{j} } \Big] \,,
   \end{split}
  \end{align}
\end{widetext}
\begin{align}
O_{ij}^{\text{spin}} &= \sum_a \frac{1}{8m_a^2} P_{a k} ( P_{a i}  S_{a (k) (j)} + P_{a j} S_{a (k) (i)} ) \,.
\end{align}
The term in Eq.\ (\ref{RS1result}) containing
$17 n_{a^{\prime}b^{\prime}}^{i} P_a{}_{j}$ actually cancels if the sums over
$a^{\prime}$ and $b^{\prime}$ are performed and may therefore be dropped.

Notice that $\Pi_{2ij}^{\rm spin}$, ${\widetilde \Pi}_{2ij}^{\rm spin}$,
and $Q_{ij}^{\prime \rm spin}$ are given by almost identical expressions,
cf.\ Equations (\ref{p2sresult}), (\ref{pi2tresult}), and (\ref{Qpresult}). This is not accidental, 
but due to similarities of their defining integrals.  With the source
mass density given by Eq.\ (\ref{H2}), we obtain from Eq.\ (\ref{PI2spin}):
\begin{equation}\label{PI2spinred}
\Pi_{2 ij}^{\text{spin}}  = - \frac{4 G}{5} \sum_a m_a \left. h^{\text{TT spin}}_{(4) i j} \right|_{\vct{x} = \hat{\vct{z}}_a} \,.
\end{equation}
Similarly, Eq.\ (\ref{PI2Tspin}) leads to
\begin{equation}\label{PI2Tspinred}
\widetilde{\Pi}_{2 ij}^{\text{spin}}  = - \frac{4 G}{5} \sum_a m_a \left. (h^{\text{TT spin}}_{(4) i j})\primevar{{}} \right|_{\vct{x} = \hat{\vct{z}}_a} \,.
\end{equation}
Notice that in this expression no regularization is needed for taking
$\vct{x} = \hat{\vct{z}}_a$, as primed and unprimed objects are not identified
yet. In contrast to that there may be contributions from Hadamard regularization
in Eq.\ (\ref{PI2spinred}). However, for the spin-dependent part, no such
contributions appear (in contrast to the nonspinning case in
\cite{Jaranowski:Schafer:1997}), which explains the great similarity between
$\Pi_{2ij}^{\rm spin}$ and ${\widetilde \Pi}_{2ij}^{\rm spin}$. Further,
insertion of Eq.\ (\ref{Npotential}) into Eq.\ (\ref{Qpspin}) leads to:
\begin{align}
Q_{ij}^{\prime \rm spin} &= - \frac{1}{4} \int \dd^3 x \, h^{\text{TT spin}}_{(4) i j} (\mathcal{H}^{\rm matter}_{(2)})\primevar{{}} \,, \nonumber \\
	&= - \frac{1}{4} \sumap m_{a^{\prime}} \left. h^{\text{TT spin}}_{(4) i j} \right|_{\vct{x} = \hat{\vct{z}}_{a^{\prime}}} \,,
\end{align}
after performing several partial integrations and using Eqs.\ (\ref{B4ij}) and
(\ref{htt4}). Here also no regularization is needed. The similarity to
Eqs.\ (\ref{PI2Tspin}) or (\ref{PI2Tspinred}) is obvious. The difference is simply an
overall factor and a mutual exchange of primed and unprimed variables.

\section{Energy loss of a binary system\label{sec:energyloss}}
\subsection{Derivation of the energy loss from the Hamiltonian}
The instantaneous (near-zone)
energy loss of a two-body system due to gravitational radiation can be written
in the form (see, e.g., \cite{Jaranowski:Schafer:1997,
Konigsdorffer:Faye:Schafer:2003}):
\begin{align}
  {\cal L}_{\le 3.5{\rm PN}}^{\rm inst} = - \dtex (H^{\rm int}_{2.5{\rm PN}}+H^{\rm int}_{3.5{\rm PN}}) \,.
\label{eq:fluxmain}
\end{align}
Notice that this energy loss is gauge-dependent in contrast to the energy flux at infinity.

We substitute Eqs.\ (\ref{eq:H25}) and (\ref{eq:H35final}) into Eq.\ (\ref{eq:fluxmain}) 
(For the point-mass part of $H_{3.5{\rm PN}}^{\rm int}$ this was already done in
\cite{Jaranowski:Schafer:1997}). After that, we need to eliminate the time
derivatives in Eq.\ (\ref{eq:fluxmain}) using the leading-order spin-orbit,
spin(1)-spin(2), and Newtonian equations of motion derived from the
corresponding Hamiltonians [see, e.g., Eqs.\ (7.28) and (7.29) in
\cite{Steinhoff:Schafer:Hergt:2008}),
\begin{subequations}
  \begin{align}
    \dot{\hat{z}}_{1^\prime}^i &= \frac{p_{1^\prime}^i}{m_{1^\prime}} -
             \frac{G}{2m_{1^\prime}}
             \frac{n^j_{1^\prime 2^\prime}}{r^2_{1^\prime 2^\prime}} 
             \left( 3 m_{2^\prime} S_{1^\prime}{}_{(j)}{}_{(i)} + 4 m_{1^\prime}
             S_{2^\prime}{}_{(j)}{}_{(i)} \right) \,, \\
    \dot{\hat{z}}_{2^\prime}^i &= (1 \rightleftharpoons 2)\,,
  \end{align}
\label{eq:xdot}
\end{subequations}
\begin{subequations}
\begin{widetext}
  \begin{align}
    \dot{p}_{1^\prime}^i &=  
        -\frac{G m_{1^\prime}m_{2^\prime}}{r^2_{1^\prime
            2^\prime}}n^i_{1^\prime 2^\prime}
         + \frac{G n_{1^\prime 2^\prime}^{i}}{r^3_{1^\prime 2^\prime}} 
        \Biggl\{ 
           \frac{9}{2}
              \frac{m_{2^\prime}}{m_{1^\prime}}{
                \bigg[\left({\mathbf p_{1^\prime} \times {\mathbf
                        S_{1^\prime}}}\right)\cdot {\mathbf
                    n}_{1^\prime 2^\prime}\bigg]}
           -6 {\bigg[\left({\mathbf p_{2^\prime} \times {\mathbf
                        S_{1^\prime}}}\right)\cdot {\mathbf
                    n}_{1^\prime 2^\prime}\bigg]}
           \nnlq \quad
           + \frac{15}{r_{1^\prime 2^\prime}} \left( {\bf n}_{1^\prime 2^\prime} \cdot
             {\bf S}_{1^\prime} \right)
             \left( {\bf n}_{1^\prime 2^\prime} \cdot
             {\bf S}_{2^\prime} \right)
           - \frac{3}{r_{1^\prime 2^\prime}} {\left( {\bf S}_{1^\prime} \cdot
             {\bf S}_{2^\prime} \right) }  
           + 6 {\bigg[\left({\mathbf p_{1^\prime} \times {\mathbf
                        S_{2^\prime}}}\right)\cdot {\mathbf
                    n}_{1^\prime 2^\prime}\bigg]}
            \nnlq \quad
           - \frac{9}{2} \frac{m_{1^\prime}}{m_{2^\prime}}
            {\bigg[\left({\mathbf p_{2^\prime} \times {\mathbf
                        S_{2^\prime}}}\right)\cdot {\mathbf
                    n}_{1^\prime 2^\prime}\bigg]}
        \Biggl\}
        \nl
        +\frac{G}{r^3_{1^\prime 2^\prime}}
      \Biggl\{
        -\frac{3}{2}\frac{ m_{2^\prime}}{m_{1^\prime}}
         \left( {\bf p}_{1^\prime} \times {{\bf S}_{1^\prime}}
         \right)^i
        +2 \left( {\bf p}_{2^\prime} \times {{\bf S}_{1^\prime}}
         \right)^i
        -\frac{3}{r_{1^\prime 2^\prime}}\left({\bf n}_{1^\prime
            2^\prime} \cdot {\bf S}_{2^\prime}  \right)S^i_{1^\prime} 
        -2 \left( {\bf p}_{1^\prime} \times {{\bf S}_{2^\prime}}
         \right)^i
          \nnlq \quad
        +\frac{3}{2}\frac{ m_{1^\prime}}{m_{2^\prime}}
         \left( {\bf p}_{2^\prime} \times {{\bf S}_{2^\prime}}
         \right)^i
        -\frac{3}{r_{1^\prime 2^\prime}}\left({\bf n}_{1^\prime
          2^\prime} \cdot {\bf S}_{1^\prime}  \right)S^i_{2^\prime} 
       \Biggl\} \nonumber \,,
  \end{align}
\end{widetext}
  \begin{align}
    \dot{p}_{2^\prime}^i &= (1 \rightleftharpoons 2)\,.
  \end{align}
\label{eq:pdot}
\end{subequations}
Note that because the 2.5PN order Hamiltonian does not have spin
contributions as we discussed in Sec.\ \ref{sec:calculation},
we do not include the 1PN point-mass terms
because substituting them into the 2.5PN Hamiltonian only produces
point-mass terms at 3.5PN order, while substituting them into the 3.5PN
Hamiltonian only produces 4.5PN terms which is beyond the scope of
this paper.

At this point, we no longer need to distinguish the difference between
the primed and unprimed variables. Using the Hadamard regularization
method, we remove the singularities produced by the limit
$\hat{\bf z}_{1^\prime} \rightarrow \hat{\bf z}_1$ and
$\hat{\bf z}_{2^\prime} \rightarrow \hat{\bf z}_2$ and obtain an expression of
the energy loss in terms of $\hat{\bf z}_{1(2)}$ and ${\bf p}_{1(2)}$. By
realizing that $\dot{\hat{\bf z}}_{a} \equiv {\bf v}_a$, we may use
\begin{subequations}\label{eq:p1PN}
\begin{align}
  p_1^i &= {m_1} v_1^i 
              -\frac{G}{2} \frac{n{}^{j}}{r^{2}} 
                 \left( 3 m_2\, S_{1(j)(i)} + 4 m_1\,
                   S_{2(j)(i)}\right)\,, \\
  p_2^i &= (1 \rightleftharpoons 2) \,,
\end{align}
\end{subequations}
to express the particle momenta ${\mathbf p}_a$ in
terms of the particle coordinate velocities ${\mathbf v}_a$,
which can be easily obtained from Eq.\ (\ref{eq:xdot}). Here, $r = r_{12}$, and ${\bf n}={\bf n}_{12}$.
Note we do not include the 1PN point-mass terms in this expression for
the reason described above.

To put the energy loss into a more convenient form, we rewrite the
individual masses $m_{1}\,, m_2$ into the total mass of the system $M
\equiv m_1 + m_2$, the reduced mass $\mu \equiv m_1 m_2/M$, and the
symmetric mass-ratio parameter $\eta \equiv \mu/M$ using the
relations (assuming $m_1 \ge m_2$):
\begin{subequations}\label{eq:mtocom}
  \begin{align}
    m_1 &= \frac{\mu}{2\eta}\left( 1 + \sqrt{1-4\eta} \right)\,, \\
    m_2 &= \frac{\mu}{2\eta}\left( 1 - \sqrt{1-4\eta} \right)\,.
  \end{align}
\end{subequations}
We also transform the individual coordinate velocities of each particle
into the center of mass frame using the relations:
\begin{widetext}
\begin{subequations}
  \begin{align}
    {{\bf v}_1} &= \frac{2\eta {\bf v}} {1 + \sqrt{1 -4 \eta}} 
              + \frac{G}{4 r^2}
               \left[ \left({\bf n}\times{{\mathbf S}_1}\right)  \left(-1 + \sqrt{1 -4
                \eta}\right)  + \left({\bf n} \times {{\mathbf S}_2}{}\right)  \left(1 + {\sqrt{1 -4
                  \eta}}  \right)\right] \,, \\
    {\bf v}_2 &= \frac{-2\eta {\bf v}} {1 - \sqrt{1 -4 \eta}} 
              + \frac{G}{4 r^2}
               \left[ \left({\bf n}\times{{\mathbf S}_1}\right)  \left(-1 + \sqrt{1 -4
                \eta}\right)  + \left({\bf n} \times {{\mathbf S}_2}{}\right)  \left(1 + {\sqrt{1 -4
                  \eta}}  \right)\right]     \,,
  \end{align}
\label{eq:vtocom}
\end{subequations}
\end{widetext}
where ${\bf v} = {\bf v}_1-{\bf v}_2$ is the relative velocity, ${\mathbf S}_a$ is the individual spin. Notice that here we do not include the 1PN
point-mass terms (see, e.g., Eq.\ (3.13) in \cite{Konigsdorffer:Faye:Schafer:2003}) because the 1PN
corrections of ${\bf v}_a$ can only produce 3.5PN terms in the flux
when substituted into ${\cal L}_{\le 2.5{\rm PN}}^{\rm inst}$, which is
independent of spins, therefore the 1PN point-mass terms in ${\bf v}_a$
do not contribute any spin-dependent terms at the formal 3.5PN order.

After eliminating the coordinate velocity ${\bf v}_1$ and ${\bf v}_2$
by means of Eq.\ (\ref{eq:vtocom}), the spin-orbit and spin(1)-spin(2) (S$_1$S$_2$)
part of the instantaneous energy loss ${\cal L}_{\le 3.5{\rm PN}}^{\rm inst}$ can
be written as:
\begin{widetext}
\begin{subequations}
  \begin{align}
    \begin{split}\label{eq:Ladm25}
      {\cal L}_{\le 2.5{\rm PN}}^{\rm inst} &= \frac{4}{15} \frac{G^2 M^{3} \eta^2}
      {r^{3}} \Biggl\{ 2\frac{G^2 M^2}{r^2} + 45 ({\mathbf n} \cdot
      {\mathbf v})^{4} -60 ({\mathbf n} \cdot {\mathbf v})^{2} v^{2}
      + 11 v^{4} + \frac{G M}{r} ( -9 ({\mathbf n} \cdot {\mathbf
        v})^{2} + 11 v^{2})\Biggl\} \,,
    \end{split}\\
{\cal L}^{\rm inst, SO}_{\le \rm 3.5PN} &=
-\frac{G^2 M^2 \eta^2}{15 r^5} \Biggl\{ 
 {\left( \hat{\mathbf L}_{\rm N} \cdot {\bf \xi} \right) }  
\bigg[  74\frac{G^2 M^2}{r^2} + 420 ({\mathbf n} \cdot
{\mathbf v})^{4} - 510 ({\mathbf n} \cdot {\mathbf v})^{2} {v}^{2} + 66
{v}^{4} 
+ \frac{G M}{r} \left( 54 ({\mathbf n} \cdot {\mathbf v})^{2}   +22{v}^{2}  \right) \bigg]  
\nnlq \quad
+ {\left( \hat{\mathbf L}_{\rm N} \cdot {\cal S} \right) }  \bigg[ 140
\frac{G^2 M^2}{r^2} +840({\mathbf n} \cdot
  {\mathbf v})^{4} -840 ({\mathbf n} \cdot {\mathbf v})^{2}
  {v}^{2} +96  v^{4}  
+ \frac{G M}{ r} \left( 336({\mathbf n} \cdot {\mathbf v})^{2}  + 84 { v}^{2}  \right) \bigg] \Biggl\}
\,, \label{eq:Lsoadm} \\
{\cal L}^{\rm inst, S_1S_2}_{\le 3.5{\rm PN}} &= 
        \frac{2}{15}\frac{G^2 M\eta}{r^5}\Biggl\{
           {\left( {{\mathbf S}_1}\cdot {{\mathbf S}_2} \right) }
          \bigg[ 12 \frac{G^2 M^{2}}{r^2} -120 ({\mathbf n} \cdot {\mathbf v})^{2}
              v^{2}  + 24 v^{4}  
         + \frac{G M}{r} \left(192 v^{2}   -348
           \left({\mathbf n} \cdot {\mathbf v}\right)^{2}  \right) \bigg]
        \nnlq \quad \quad
        +   {\left( {{\mathbf S}_1}\cdot {\bf v} \right) }  
           {\left({{\mathbf S}_2}\cdot {\bf v} \right) }  
          \left[ 184 \frac{G M}{r}  
          -450 ({\mathbf n} \cdot {\mathbf v})^{2} + 138 v^{2}  \right]  
        \nnlq \quad \quad
        + \left[ ({\mathbf n} \cdot {\mathbf v}) 
          {\left( {\bf n}\cdot {{\mathbf S}_2} \right) }  
          {\left( {{\mathbf S}_1}\cdot {\bf v} \right) }
          + ({\mathbf n} \cdot {\mathbf v}) 
          {\left( {\bf n} \cdot {{\mathbf S}_1} \right) }  
          {\left( {{\mathbf S}_2}\cdot {\bf v} \right) } 
        \right]  
          \left[ - 546 \frac{G M}{r}   +  1785 ({\mathbf n} \cdot
            {\mathbf v})^{2} - 1005 v^{2}    \right]
        \label{eq:Lssadm} \nlq \quad \quad
        + 
         {\left( {\bf n}\cdot {{\mathbf S}_1} \right) }  
         {\left( {\bf n}\cdot {{\mathbf S}_2} \right) }  
         \bigg[ -36 \frac{G^2 M^2}{r^2}
           -5670 ({\mathbf n} \cdot {\mathbf v})^{4}+4620
         ({\mathbf n} \cdot {\mathbf v})^{2} v^{2}  
         -390v^{4} 
        \nnlq \quad \quad \quad \quad
        + \frac{G M}{ r} \left( - 600 v^{2}  +
        1536 ({\mathbf n} \cdot {\mathbf v})^{2} \right) \bigg]
        \Biggl\}\,, \nonumber
\end{align}
\label{eq:Ladm}
\end{subequations}
\end{widetext}
with $v = |\vct{v}|$, 
${\cal S} \equiv {\bf S}_1 + {\bf S}_2$,
${\bf \xi} \equiv (m_2/m_1){\bf S}_1+ (m_1/m_2){\bf S}_2$ are the spin variables, and 
$\hat{\bf L}_N \equiv r {\bf n} \times {\bf v}$ is the Newtonian
orbital angular momentum per reduced mass.

\subsection{Comparison with other results}
References \cite{Will:2005, Wang:Will:2007} recently computed, using the method
of \emph{direct integration of the relaxed Einstein equations}
\cite{Pati:Will:2000,Pati:Will:2002}, the leading-order spin-orbit and
spin(1)-spin(2) equations of motion and the corresponding energy loss in
\emph{harmonic coordinates}. In this subsection, we shall prove that our result
is actually equivalent to the results in \cite{Will:2005, Wang:Will:2007}.

In order to compare the instantaneous energy loss, we first need to find the
transformation between our ADM canonical variables
$(\hat{\bf z}_a,{\bf v}_a\equiv \dot{\hat{\bf z}}_a,{\bf S}_a)$ and the
``harmonic coordinate'' variables
$({\bf y}_a,{\bf V}_a\equiv \dot{\bf y}_a, {\bf S}_a^{\rm WW})$. Because
the quantity we are comparing is the energy loss
${\cal L}_{\le 3.5{\rm PN}}^{\rm inst}$ at formal 3.5PN order, which is only one
formal order higher than the leading-order energy loss ${\cal L}_{\le 2.5{\rm PN}}^{\rm inst}$ caused by the quadrupole
radiation of point-masses, the
coordinate transformation we are looking for only needs to be accurate up to
formal 1PN order.

It is well known that for the point-mass case the ADM coordinates are equivalent
to the harmonic coordinates at 1PN order in that they result in identical
equations of motion.
In addition, the spin-dependent part of the formal 1PN accurate transformation
$\hat{\bf z}_a({\bf y}_a,{\bf V}_a,{\bf S}^{\rm WW}_a)$ can be derived from the
well-known transformation between different \emph{spin supplementary conditions}
(SSC) (for details, see, e.g., \cite{Will:2005}). Namely, for a specific SSC
parameter $k$, which is used to fix the center of mass of the
particle, we impose the condition:
\begin{align}
  S_a^{i0} - k S_a^{ij}v_a^j =0 \,,
\end{align}
where $k$ typically has the value 1, $1 / 2$, or 0. The relation
between the center of mass for each value of $k$ can be written as:
\begin{align}
  (x_a^i)^{(k^\prime)} = (x_a^i)^{(k)} + \frac{k-k^\prime}{m_A}
S_a^{ij}(v_a^j)^{(k)} \,.
\label{eq:kssc}
\end{align}

It is straightforward to show that at formal 1PN order the SSC in our
calculation leads to $k=1/2$, which is identical to the one used in
references \cite{Will:2005, Wang:Will:2007}. Therefore, we have:
\begin{align}
  \hat{\bf z}_a({\bf y}_a,{\bf V}_a,{\bf S}^{\rm WW}_a) &= {\bf y}_a \,, \label{eq:xadm2h} \\
  {\bf v}_a({\bf y}_a,{\bf V}_a,{\bf S}^{\rm WW}_a) &\equiv \dot{\hat{\bf z}}_a  = {\bf V}_a \,. \label{eq:vadm2h}
\end{align}

Reference \cite{Damour:Jaranowski:Schafer:2008:1} has shown that the difference
between the spin parameters ${\bf S}_a$ used in the ADM formalism and the ones
used in the harmonic coordinates calculations is of formal 2PN order. In other
words, the transformation
\begin{align}
    {\bf S}_a({\bf y}_a,{\bf V}_a,{\bf S}^{\rm WW}_a) &= {\bf S}^{\rm WW}_a \,, \label{eq:sadm2h}
\end{align}
can be used in this paper.

From Eqs.\ (\ref{eq:xadm2h}) -- (\ref{eq:sadm2h}) we know that our ADM canonical
variables are actually equivalent to the harmonic gauge ones at the considered
PN order. Now we are not comparing with the harmonic gauge energy loss
given in \cite{Pati:Will:2002, Will:2005, Wang:Will:2007} directly, but with the
far-zone energy flux, which was shown to agree with the former (up to an nonphysical total time derivative). When comparing our result
Eqs.\ (\ref{eq:Ladm}) to the far-zone flux
$\left[{\cal L}_{\le 3.5{\rm PN}}\right]^{\text{far-zone}}$, for the purpose of
this paper, only the parts
\begin{subequations}
  \begin{align}
\begin{split}
  \left[ {\cal L}_{\le 3.5{\rm PN}}\right]^{\rm inst}&= 
     [{\cal L}_{\le 2.5{\rm PN}}
    + {\cal L}^{\rm SO}_{\le 3.5{\rm PN}} \nlq
    + {\cal L}^{\rm S_1S_2}_{\le 3.5{\rm PN}}]^{\rm inst} \,,
  \label{eq:Ladmformal} 
\end{split}\\
\begin{split}
    \left[{\cal L}_{\le 3.5{\rm PN}}\right]^{\text{far-zone}}&= 
     [{\cal L}_{\le 2.5{\rm PN}}
    + {\cal L}^{\rm SO}_{\le 3.5{\rm PN}} \nlq
    + {\cal L}^{\rm S_1S_2}_{\le 3.5{\rm PN}} ]^{\text{far-zone}}
    \,,
\label{eq:Lharmonicformal}
\end{split}
\end{align}
\end{subequations}
are relevant to this paper, where for the instantaneous energy loss in ADM coordinates we
substitute Eq.\ (\ref{eq:Ladm}) and for the far-zone flux we substitute the
expressions computed in \cite{Kidder:1995} in harmonic gauge,
\begin{subequations}
  \begin{align}
    \left[{\cal L}_{\le 2.5{\rm PN}}\right]^{\text{far-zone}} &=\frac{8}{15} \frac{G^3 M^{4}\eta^2}{ r^{4}} \left[ ( -11 ({\mathbf n} \cdot {\mathbf v})^{2} + 12 v^{2}\right] , 
  \end{align}
\begin{widetext}
  \begin{align}
  \begin{split}
  \left[{\cal L}^{\rm SO}_{\le 3.5{\rm PN}}\right]^{\text{far-zone}} &=
    \frac{8}{15} \frac{G^3 M^{3}\eta^2}{ r^{6}} \Biggl\{ 
    {\left( {\hat{\mathbf L}_{\rm N}}{}\cdot {\cal \xi}{} \right) }  
          \left[ -8  \frac{G M}{r} + 18 ({\mathbf n} \cdot {\mathbf
              v})^{2} -19 v^{2}\right]  
     \nlq \quad \quad
    +   {\left( {\hat{\mathbf
          L}_{\rm N}}\cdot {\cal S}{}\right) }  
          \left[ -12 \frac{G M}{r} + 27 ({\mathbf n} \cdot {\mathbf v})^{2} -37
          v^{2}\right]  
        \Biggl\}\,,
  \end{split}\\
\begin{split}   
    \left[{\cal L}^{\rm S_1S_2}_{\le 3.5{\rm PN}}\right]^{\text{far-zone}} &= 
        \frac{4}{15} \frac{G^3 M^{2}\eta}{ r^{6}} \Biggl\{ 
          -171 ({\mathbf n} \cdot {\mathbf v}){\left( {\bf n}{}\cdot {{\bf S}_2} \right) }  {\left(
              {{\bf S}_1}\cdot {\bf v} \right) }  
          -171 ({\mathbf n} \cdot {\mathbf v}){\left( {\bf n}\cdot {{\bf S}_1} \right) }  {\left(
              {{\bf S}_2}\cdot {\bf v} \right) }  
          + 71 {\left( {{\bf S}_1}\cdot {\bf v} \right) }  {\left(
              {{\bf S}_2}\cdot {\bf v} \right) }  
           \nlq
          + {\left( {\bf n} \cdot {\bf S}_1 \right) }  
            {\left( {\bf n} \cdot{{\mathbf S}_2} \right) }  
            \left[ 807 \left({\mathbf n} \cdot {\mathbf v}\right)^{2} -504 v^{2}\right]
          + {\left( {{\mathbf S}_1}\cdot {{\mathbf S}_2} \right) }  
            \left[ -165 ({\mathbf n} \cdot {\mathbf v})^{2} + 141
              v^{2}\right] \Biggl\}\,.
\end{split}
  \end{align}
\end{widetext}
\end{subequations}
It should be noted that the sources (on the right-hand side of these equations)
are evaluated at the retarded time with respect to the flux (on the left-hand
side), which is not explicitly denoted here.
In contrast to the instantaneous near-zone energy loss, these results are actually
gauge-independent at the considered PN order, i.e., one gets exactly the same
result from Eq.\ (6.22) in \cite{Steinhoff:Wang:2009} within the ADM gauge. We
already showed this in \cite{Steinhoff:Wang:2009} for
$\left[{\cal L}^{\rm SO}_{\le 3.5{\rm PN}}\right]^{\text{far-zone}}$, and we
confirmed this for $\left[{\cal L}_{\le 2.5{\rm PN}}\right]^{\text{far-zone}}$ and
$\left[{\cal L}^{\rm S_1S_2}_{\le 3.5{\rm PN}}\right]^{\text{far-zone}}$, too.

It has been shown in \cite{Konigsdorffer:Faye:Schafer:2003} that the
spin-\emph{independent} result
$\left[{\cal L}_{\le 2.5{\rm PN}}\right]^{\rm inst} + \left[{\cal L}^{\rm PM}_{3.5{\rm PN}}\right]^{\rm inst}$ 
and the spin-orbit part of formal 3.5PN order $\left[{\cal L}^{\rm SO}_{3.5{\rm PN}}\right]^{\rm inst}$
agree with the results computed in harmonic coordinates up to a total time
derivative, which is a pure gauge effect and vanishes after orbital average
(see, e.g, \cite{Zeng:Will:2007} and \cite{Iyer:Will:1993}).

For spin-dependent instantaneous energy loss, it is possible to write the difference between Eqs.\ (\ref{eq:Ladmformal}) and
(\ref{eq:Lharmonicformal}) as a total time derivative using the identities in
Appendix \ref{sec:timediff}, which has already been presented in Appendix F of
\cite{Will:2005} and Appendix A of \cite{Wang:Will:2007}.
Taking into account the leading-order point-mass and spin contributions it holds that
\begin{multline}\label{eq:defE}
\left[ {\cal L}_{\le 2.5{\rm PN}} + {\cal L}^{\rm SO}_{\le 3.5{\rm PN}} + {\cal L}^{\rm S_1S_2}_{\le 3.5{\rm PN}} \right]^{\rm inst} \\
- \left[ {\cal L}_{\le 2.5{\rm PN}} + {\cal L}^{\rm SO}_{\le 3.5{\rm PN}} + {\cal L}^{\rm S_1S_2}_{\le 3.5{\rm PN}} \right]^{\text{far-zone}} \\
= \frac{d}{dt} \left[ E_{2.5{\rm PN}} + E^{\rm SO}_{3.5{\rm PN}} + E^{\rm S_1S_2}_{3.5{\rm PN}} \right] \,,
\end{multline}
where
\begin{subequations}
\begin{equation}\label{eq:diff25}
E_{2.5{\rm PN}} = \frac{G^2 M^3 \eta^2}{r^2} ({\mathbf n}\cdot{\mathbf v}) \bigg[ \frac{44}{15} v^2
	- \frac{12}{5} ({\mathbf n}\cdot{\mathbf v})^2 - \frac{8}{15} \frac{G M}{r} \bigg] , 
\end{equation}
\begin{widetext}
\begin{align}
\begin{split}\label{eq:diff35SO}
E^{\rm SO}_{3.5{\rm PN}}&=\frac{G^2 M^2\eta^2}{r^4}({\mathbf n}\cdot{\mathbf v})\biggl[({\hat{\mathbf L}_{\rm N}}\cdot{\cal S})\biggl(\frac{28}{15}\frac{G M}{r}+8({\mathbf n}\cdot{\mathbf v})^2-\frac{32}{15}v^2\biggr)
\nlq
+({\hat{\mathbf L}_{\rm N}}\cdot{\cal\xi})\biggl(-\frac{2}{15}\frac{G M}{r}+4({\mathbf n}\cdot{\mathbf v})^2-\frac{22}{5}v^2\biggr)\biggr]\,,
\end{split}\\
E^{\rm S_1S_2}_{3.5{\rm PN}}&=\frac{G \eta}{r^3}\biggl[({\bf n}\cdot{\mathbf S}_1)({\bf n}\cdot{\mathbf S}_2)({\mathbf n}\cdot{\mathbf v})\frac{G M}{r}\biggl(84({\mathbf n}\cdot{\mathbf v})^2-52v^2-\frac{44}{5}\frac{G M}{r}\biggr)
\nnlq
+\frac{G M}{r}\biggl(({\bf n}\cdot{\bf S}_1)({\bf S}_2\cdot{\bf v})+({\bf n}\cdot{\bf S}_2)({\bf S}_1\cdot{\bf v})\biggr)\biggl(-22({\mathbf n}\cdot{\mathbf v})^2+\frac{38}{5}v^2+\frac{22}{5}\frac{G M}{r}\biggr)
\label{eq:diff35SS} \nlq
+\frac{16}{5}\frac{G M}{r}({\bf S}_1\cdot{\bf v})({\bf S}_2\cdot{\bf v})({\mathbf n}\cdot{\mathbf v})+\frac{16}{5}\frac{G M}{r}v^2({\mathbf n}\cdot{\mathbf v})({\mathbf S}_1\cdot{\mathbf S}_2)
\biggr]\,. \nonumber
\end{align}
\end{widetext}
\end{subequations}
Note that even though the energy loss at 2.5PN order is spin-\emph{independent},
it \emph{does} need to be taken into account when comparing the
spin-\emph{dependent} energy losses because of the spin-dependent terms in
Eqs.\ \eqref{dt1} and \eqref{dt2}, which are of formal 1PN order.

It should be noted that the total time derivative on the right-hand side of
Eq.\ (\ref{eq:defE}) vanishes to the order in question when averaged over time. This means that
the time average of near-zone energy loss and far-zone energy flux agree.
Equations (\ref{eq:diff25}) -- (\ref{eq:diff35SS}) should thus be interpreted as
(gauge-dependent) energies that temporarily leave the near-zone, but never
reach the far-zone and instead move back into the near-zone at a later time.
Therefore, they have in average no effect on the near-zone energy loss.

\section{Conclusions and outlook\label{sec:conclusions}}
Based on developments in \cite{Steinhoff:Wang:2009} the leading-order
PN spin-orbit and spin(1)-spin(2) \emph{radiation-reaction}
Hamiltonians were calculated. Corresponding equations of motion were already
derived for the binary case in \cite{Will:2005, Wang:Will:2007, Zeng:Will:2007}.
The Hamiltonians given in the present paper are even valid for arbitrary many
spinning compact objects and present the dynamics in a compact form. The
derivation was performed within the ADM canonical formalism
\cite{Arnowitt:Deser:Misner:1962, *Arnowitt:Deser:Misner:2008}, which was
extended from point-masses to linear order in the single spin of the objects in
\cite{Steinhoff:Schafer:2009:2, Steinhoff:2011, Steinhoff:Schafer:Hergt:2008,
Steinhoff:Wang:2009}. The calculation of the needed integrals and their
regularization is analogous to calculations for nonspinning objects within the
ADM formalism (see, e.g., \cite{Jaranowski:1997, Jaranowski:Schafer:1997,
Jaranowski:Schafer:1998}). In particular, we applied the Hadamard finite part and
Riesz-formula based regularizations in the present paper (for the latter see
also \cite{Riesz:1949, *Riesz:1949:err}). Some integrals were checked using
Riesz kernels in arbitrary dimensions (see also
\cite{Damour:Jaranowski:Schafer:2008:2}).

The leading-order spin-orbit and spin(1)-spin(2) energy loss was computed in the
present paper from the explicit time derivative of the interaction Hamiltonian.
This was compared to well-known results for the corresponding energy flux
\cite{Kidder:1995} as a check [In \cite{Steinhoff:Wang:2009}, the leading-order
spin-orbit energy flux was rederived from the wave equation (\ref{wave})]. This
also proofs agreement with the energy loss obtained in the harmonic gauge
\cite{Pati:Will:2002, Will:2005, Wang:Will:2007} and thus provides an important
check of the ADM canonical formalism for \emph{spinning} objects, which was
derived only very recently \cite{Steinhoff:Schafer:2009:2, Steinhoff:2011,
Steinhoff:Schafer:Hergt:2008, Steinhoff:Wang:2009}. Notice that the interaction
Hamiltonian in the form of Eq.\ (\ref{Hint}) also gives essential contributions to the
next-to-next-to-leading-order conservative Hamiltonians
\cite{Hartung:Steinhoff:2011:1, Hartung:Steinhoff:2011:2}.

The spin-orbit radiation-reaction Hamiltonian derived in the present paper,
which is at 3.5PN when counted in a formal way, is actually of the order 4PN for
maximally rotating objects (see also Appendix A of \cite{Steinhoff:Wang:2009}).
A derivation of all spin-dependent 4PN Hamiltonians for maximally rotating
objects should be envisaged in the future. The most complicated Hamiltonian at
this level is the conservative next-to-next-to-leading-order spin(1)-spin(2)
one, but it has already been derived very recently (see
\cite{Hartung:Steinhoff:2011:2}, and also \cite{Levi:2011} for a corresponding
potential). Notice that all Hamiltonians for maximally rotating black holes are
known to 3.5PN order \cite{Hartung:Steinhoff:2011:1}.

Further, the leading-order spin-orbit and spin(1)-spin(2) radiation-reaction
equations of motion can be obtained from the Hamiltonians derived in the present
paper and compared with the results from \cite{Will:2005, Wang:Will:2007,
Zeng:Will:2007} in the future. Primed and unprimed variables must be identified
in the equations of motion, which requires further application of regularization
techniques. Finally, one may transform the general equations of motion into
secular equations of motion for the orbital elements, which has already been derived 
in \cite{Gergely:Perjes:Vasuth:1998, Gergely:1999, Gergely:2000} using energy
and angular momentum balance.

\acknowledgments
We thank P.\ Jaranowski for sharing his insight in the
calculation of the 3.5PN point-mass Hamiltonian.
J.S.\ is further grateful to M.\ Tessmer for useful discussions.
This work is supported by the Deutsche Forschungsgemeinschaft (DFG) through
SFB/TR7 ``Gravitational Wave Astronomy,'' project STE 2017/1-1, and GRK 1523,
and by the FCT (Portugal) through PTDC project CTEAST/098034/2008.

\appendix

\section{Total time derivatives\label{sec:timediff}}
The identities for total time derivatives needed to compare the instantaneous
near-zone energy loss and the far-zone energy flux were already provided in
Appendix F of \cite{Will:2005} and Appendix A of \cite{Wang:Will:2007}. They
read\footnote{There was a misprint in Eq.\ (A1) of \cite{Wang:Will:2007}, we made
appropriate changes in the expression here and marked its position by
\fbox{\rule[0.1cm]{0cm}{0.1cm}$\ldots$}.}:
\begin{widetext}
\begin{subequations}
  \begin{align}
      \frac{d}{dt} \left ( \frac{v^{2s} {\dot r}^p}{r^q} \right ) &=
\frac{v^{2s-2} {\dot r}^{p-1}}{r^{q+1}}
\Biggl\{ pv^4 - (p+q)v^2{\dot r}^2 - 2s{\dot r}^2\frac{G M}{r}
-pv^2\frac{G M}{r} + \frac{p}{2}\frac{G v^2}{r^3} {\bf {\hat L}}_{\rm N} \cdot
(4 {\bf {\cal S}} + 3 {\cal \xi})
\nnl
-6s \frac{G \dot r}{\mu r^3} \left [ {\dot r}\left({{\bf S}_1\cdot{\bf
        S}_2}\right) + \left({\bf v}\cdot
    {\bf S}_1\right)\left({\bf n}\cdot{\bf S}_2\right)
  + \left({\bf v}\cdot
    {\bf S}_1\right)\left({\bf n}\cdot{\bf S}_2\right)
  -5{\dot r}\left({\bf n}\cdot{\bf S}_1\right)\left({\bf n}\cdot{\bf S}_2\right)\right]
\label{dt1} \nl
- 3p\frac{G v^2}{\mu r^3} \left [ \left({{\bf S}_1\cdot{\bf S}_2}\right)
  - 3\left({\bf n}\cdot{{\bf
      S}_1}\right)\left({\bf n}\cdot{{\bf S}_2}\right) \right ]
\Biggl\}
\,, \nonumber \\
      \frac{d}{dt} \left ( \frac{v^{2s} {\dot r}^p}{r^q}{\bf {\hat
            L}}_{\rm N} \right ) &=
            \frac{v^{2s-2} {\dot r}^{p-1}}{r^{q+1}}\Biggl\{
            \left[  pv^4 - (p+q)v^2{\dot r}^2 - 2s{\dot r}^2\frac{G M}{r}
-pv^2\frac{G M}{r} \right]{\bf {\hat L}}_{\rm N}
           \nnl
           +  \left( \frac{p}{2}\frac{G v^2}{r^3}{\bf {\hat
            L}}_{\rm N} \cdot (4 {\bf {\cal S}} + 3 {\cal \xi})\right) {\bf
          {\hat L}}_{\rm N} 
      -\frac{G v^2 {\dot r}}{r}{\mathbf n} \times\left(
        ({\mathbf v} - \frac{3}{2}{\dot r}{\mathbf n})\times (4 {\bf
          {\cal S}} + 3 {\cal \xi})\right)  
           \label{dt2} \nl
             - 6 s \frac{G \dot r}{\mu r^3} \Big[
                 {\dot r} \left( {\bf S}_1 \cdot {\bf S}_2 \right) +
                 ({\bf v}\cdot{\bf S}_1)({\bf n}\cdot{\bf S}_2) +
                 ({\bf v}\cdot{\bf S}_2)({\bf n}\cdot{\bf S}_1)
                - 5 {\dot r} ({\bf n}\cdot{\bf S}_1)({\bf n}\cdot{\bf S}_2)
              \Big]
          {\bf {\hat L}}_{\rm N}
           \nnl
          - 3 p \frac{G v^2}{\mu r^3} \left[ \left( {\bf S}_1 \cdot
             {\bf S}_2 \right) - 3 ({\bf n}\cdot{\bf
             S}_1)({\bf n}\cdot{\bf S}_2) \right] {\bf {\hat L}}_{\rm N}
          - {\framebox 3} \frac{G v^2 {\dot r}}{\mu r^2}\left[ ({\bf n}\times{\bf
             S}_1)({\bf n}\cdot{\bf S}_2)+({\bf n}\times{\bf
             S}_2)({\bf n}\cdot{\bf S}_1) \right]
            \Biggl\} \,, \nonumber \\
\frac{d}{dt} \left ( \frac{v^{2s} {\dot r}^p}{r^q} x^ix^j \right ) &=
\frac{v^{2s-2} {\dot r}^{p-1}}{r^{q+1}}
\left \{  \left [pv^4 - (p+q)v^2{\dot r}^2 - 2s{\dot r}^2\frac{G M}{r}
-pv^2\frac{G M}{r} \right ] x^ix^j
+2 v^2{\dot r} r x^{(i}v^{j)} \right \} \,,
\label{dt3} \\
\frac{d}{dt} \left ( \frac{v^{2s} {\dot r}^p}{r^q} v^iv^j \right ) &=
\frac{v^{2s-2} {\dot r}^{p-1}}{r^{q+1}}
\left \{  \left [pv^4 - (p+q)v^2{\dot r}^2 - 2s{\dot r}^2\frac{G M}{r}
-pv^2\frac{G M}{r} \right ] v^iv^j
-2 G M \frac{v^2{\dot r}}{r^2} x^{(i}v^{j)} \right \} \,,
\label{dt4}\\
\frac{d}{dt} \left ( \frac{v^{2s} {\dot r}^p}{r^q} x^iv^j \right ) &=
\frac{v^{2s-2} {\dot r}^{p-1}}{r^{q+1}}
\left \{  \left [pv^4 - (p+q)v^2{\dot r}^2 - 2s{\dot r}^2\frac{G M}{r}
-pv^2\frac{G M}{r} \right ] x^iv^j
+ v^2{\dot r} r \left ( v^iv^j - \frac{G M}{r} n^in^j \right ) \right \} \,,
\label{dt5}
\end{align}
\end{subequations}
where $\dot r \equiv ({\bf n} \cdot {\bf v})$, and $s\,,p\,,q$ are
non-negative integers.
\end{widetext}




\ifnotprd
\bibliographystyle{utphys}
\fi

\ifarxiv
\providecommand{\href}[2]{#2}\begingroup\raggedright\endgroup

\else
\bibliography{spin_rr_adm}
\fi

\end{document}